\documentclass[twocolumn]{aastex631}

\usepackage{graphicx}% Include figure files
\usepackage{dcolumn}% Align table columns on decimal point
\usepackage{bm}% bold math
\usepackage{subfigure}
\usepackage[nointegrals]{wasysym}
\usepackage{verbatim}
\usepackage{color}
\usepackage{amsmath}
\usepackage[utf8]{inputenc}

\newcommand{ \omchmin } { 0.005 } 
\newcommand{ \omchmax } { 0.99 } 
\newcommand{ \ombhmin } { 0.005 } 
\newcommand{ \ombhmax } { 0.1 } 
\newcommand{ \logamin } { 1.61 } 
\newcommand{ \logamax } { 4.0 }

\newcommand{ \nsmin } { 0.8 } 
\newcommand{ \nsmax } { 1.2 }

\newcommand{ \okmin } { -0.3 } 
\newcommand{ \okmax } { 0.3 } 
\newcommand{ \mnumin } { 0.0 } 
\newcommand{ \mnumax } { 5.0 }

\newcommand{ \obmean } { 0.02233 } 
\newcommand{ \obsigma } { 0.00036 } 
\newcommand{ \nsmean } { 0.96 } 
\newcommand{ \nssigma } { 0.02 } 
\newcommand{ \taumin } { 0.01 } 
\newcommand{ \taumax } { 0.8 } 
\newcommand{ \thetamin } { 0.5 } 
\newcommand{ \thetamax } { 10 } 
\newcommand{ \hmin } { 40 } 
\newcommand{ \hmax } { 100 }

\newcommand{\loga}{\mathrm{ln}(10^{10}A_{\mathrm{s}})}
\newcommand{\thetamc}{100\theta_{\rm MC}}
\newcommand{\omch}{\Omega_{\mathrm{c}} h^2}
\newcommand{\ombh}{\Omega_{\mathrm{b}} h^2}
\newcommand{\ns}{n_{\mathrm{s}}}
\newcommand{\mnu}{\sum m_{\nu}}
\newcommand{\LCDM}{$\Lambda\rm{CDM}$}
\newcommand{\Planck}{{\it{Planck}}}
\newcommand{\WMAP}{{\it{WMAP}}}

\newcommand{\kcmb}{\kappa_\mathrm{CMB}}

\newcommand{\si}[1]{{\rm{#1}}}

\newcommand{\omegam}{\Omega_{\rm m}}
\newcommand{\omegak}{\Omega_{k}}
\newcommand{\omegal}{\Omega_{\Lambda}}
\newcommand{\seightg}{\sigma_8(\omegam/0.3)^{0.5}}
\newcommand{\seightc}{\sigma_8(\omegam/0.3)^{0.25}}
\newcommand{\scmbl}{S_8^{\mathrm{CMBL}}}
\newcommand{\NPIPE}{\texttt{NPIPE}}

\newcommand{\fits}{\texttt{FITS}}
\newcommand{\threecrosstwo}{3\times 2\,\text{pt}}

\newcommand{\aseightprecision}{1.8}
\newcommand{\aseightmean}{0.819}
\newcommand{\aseighterr}{0.015}
\newcommand{\ahprecision}{1.6}
\newcommand{\ahmean}{68.3}
\newcommand{\aherr}{1.1}
\newcommand{\ahmeanalt}{65.0}
\newcommand{\aherralt}{3.2}
\newcommand{\amnu}{0.13}
\newcommand{\amnus}{0.13}

\newcommand{\apseightprecision}{1.6}
\newcommand{\apseightmean}{0.812}
\newcommand{\apseighterr}{0.013}
\newcommand{\aphmean}{68.1}
\newcommand{\apherr}{1.0}
\newcommand{\aphmeanalt}{64.9}
\newcommand{\apherralt}{2.8}
\newcommand{\apmnu}{0.13}
\newcommand{\apmnuh}{0.16}

\usepackage{aas_macros}

\begin{document}
\setcounter{tocdepth}{1}

\title{The Atacama Cosmology Telescope: DR6 Gravitational Lensing Map and Cosmological Parameters}
 \shorttitle{ACT DR6 Lensing Map and Cosmology}
  \shortauthors{Madhavacheril, Qu, Sherwin, MacCrann, Li et al.}

%%%%%%%%%%%%%%%%%%%%%%%%%%%%%%%%%%%%%%%%%%%%%%%%%%%%%%%%%%%%%%%%%%%%%%%%%%%
% WARNING: This LaTeX block was generated automatically by authors_new.py
% Do not change by hand: your changes will be lost.

\author{Mathew~S.~Madhavacheril}\affiliation{Department of Physics and Astronomy, University of
Pennsylvania, 209 South 33rd Street, Philadelphia, PA, USA 19104}\affiliation{Perimeter Institute for Theoretical Physics, Waterloo, Ontario, N2L 2Y5, Canada}
\author{Frank~J.~Qu}\affiliation{DAMTP, Centre for Mathematical Sciences, University of Cambridge, Wilberforce Road, Cambridge CB3 OWA, UK}
\author{Blake~D.~Sherwin}\affiliation{DAMTP, Centre for Mathematical Sciences, University of Cambridge, Wilberforce Road, Cambridge CB3 OWA, UK}\affiliation{Kavli Institute for Cosmology Cambridge, Madingley Road, Cambridge CB3 0HA, UK}
\author{Niall~MacCrann}\affiliation{DAMTP, Centre for Mathematical Sciences, University of Cambridge, Wilberforce Road, Cambridge CB3 OWA, UK}
\author{Yaqiong~Li}\affiliation{Department of Physics, Cornell University, Ithaca, NY, USA 14853}
\author{Irene~Abril-Cabezas}\affiliation{DAMTP, Centre for Mathematical Sciences, University of Cambridge, Wilberforce Road, Cambridge CB3 OWA, UK}
\author{Peter~A.~R.~Ade}\affiliation{School of Physics and Astronomy, Cardiff University, The Parade, 
Cardiff, Wales, UK CF24 3AA}
\author{Simone~Aiola}\affiliation{Center for Computational Astrophysics, Flatiron Institute, 162 5th Avenue, New York, NY 10010 USA}\affiliation{Joseph Henry Laboratories of Physics, Jadwin Hall,
Princeton University, Princeton, NJ, USA 08544}
\author{Tommy~Alford}\affiliation{Department of Physics, University of Chicago, Chicago, IL 60637, USA}
\author{Mandana~Amiri}\affiliation{Department of Physics and Astronomy, University of
British Columbia, Vancouver, BC, Canada V6T 1Z4}
% ...

% End auto-generated block
%%%%%%%%%%%%%%%%%%%%%%%%%%%%%%%%%%%%%%%%%%%%%%%%%%%%%%%%%%%%%%%%%%%%%%%%%%%

\begin{abstract}
We present cosmological constraints from a gravitational lensing mass map covering $9400\,\si{deg}^2$ reconstructed from measurements of the cosmic microwave background (CMB) made by the Atacama Cosmology Telescope (ACT) from 2017 to 2021. In combination with measurements of baryon acoustic oscillations (BAO, from SDSS and 6dF) and big bang nucleosynthesis (BBN), we obtain the amplitude of matter fluctuations $\sigma_8 = \aseightmean \pm \aseighterr$ at \aseightprecision\% precision, $S_8\equiv\seightg=0.840\pm0.028$ and the Hubble constant $H_0= (\ahmean \pm \aherr)\, \text{km}\,\text{s}^{-1}\,\text{Mpc}^{-1}$ at \ahprecision\% precision. A joint constraint with CMB lensing measured by the \Planck\ satellite yields even more precise values: $\sigma_8 = \apseightmean \pm \apseighterr$, $S_8\equiv\seightg=0.831\pm0.023$ and $H_0= (\aphmean \pm \apherr)\, \text{km}\,\text{s}^{-1}\,\text{Mpc}^{-1}$. These measurements are in excellent agreement with $\Lambda$CDM-model extrapolations from the CMB anisotropies measured by \Planck. To compare these ACT CMB lensing constraints to those from the KiDS, DES, and HSC galaxy surveys, we revisit those data sets with a uniform set of assumptions, and find $S_8$ from all three surveys are lower than that from ACT+\Planck~lensing by varying levels ranging from 1.7--2.1$\sigma$. These results motivate further measurements and comparison, not just between the CMB anisotropies and galaxy lensing, but also between CMB lensing probing $z\sim 0.5-5$ on mostly-linear scales and galaxy lensing at $z\sim 0.5$ on smaller scales.  We combine our CMB lensing measurements with CMB anisotropy data to constrain extensions of \LCDM, limiting the sum of the neutrino masses to $\sum m_{\nu} < \amnus$\,eV (95\% c.l.), for example. We also provide constraints independent of the CMB anisotropy data from \Planck, using instead \WMAP\ and ACT CMB data to constrain the spatial curvature density to $-0.016 < \omegak <  0.012 ~({\rm 95\%~ c.l.})$ and the dark energy density to $\omegal = 0.68\pm 0.01$. We describe the mass map and related data products that will enable a wide array of cross-correlation science. This map is signal-dominated for lensing mass map multipoles $L<150$ with the large scales measured with a signal-to-noise ratio around twice that of \Planck.  Our results provide independent confirmation that the universe is spatially flat, conforms with general relativity, and is described remarkably well by the \LCDM~model, while paving a promising path for neutrino physics with gravitational lensing from upcoming ground-based CMB surveys.

\end{abstract}

\section{Introduction}

The cosmic microwave background (CMB) provides a view of the early universe ($z\gtrsim 1100$ or age $t\lesssim 375,000$ years) through primary anisotropies in the relic radiation left over from the hot big bang. Later, as the universe became transparent after recombination, expanded, and cooled, CMB photons continued to experience occasional interactions with structures forming over cosmic time under the influence of gravity.  These interactions left behind secondary imprints in the CMB providing a window into the late-time universe: a view of large-scale structure complementary to galaxy and intensity-mapping surveys. In particular, CMB photons travel through all the mass in
the observable universe as it develops into large-scale structures; the ensuing gravitational deflections manifest as distortions on arcminute-scales in the CMB that retain coherence over degree scales, the latter corresponding to the size of typical lenses projected along the line-of-sight ($\sim 300$ Mpc). The lensing distortions are distinguished from the Gaussian and statistically isotropic fluctuations in the CMB through the use of quadratic estimators \citep{HuOk}, resulting in comprehensive mass maps, dominated by dark matter, and probing primarily linear scales. (See \citealt{0601594} for a review.) 

Precise measurements of the CMB on small scales have already allowed the extraction of this secondary lensing signal (probing the late-time universe) from underneath the primary CMB information (probing the early universe). CMB lensing measurements to date include those from the \WMAP\ satellite  \citep{0705.3980}, from ground-based surveys including the Atacama Cosmology Telescope (ACT; \citealp{1103.2124,1611.09753}), the South Pole Telescope (SPT; e.g., \citealp{1202.0546,1910.07157,2012.01709}), BICEP2/Keck Array \citep{1606.01968} and POLARBEAR \citep{1312.6646,1911.10980}, and from the \Planck\ satellite~\citep{1303.5077,1502.01591,1807.06210,2206.07773}.

While a standard cosmological model has emerged based on precise measurements of the primary CMB anisotropy over the last few decades, it is currently undergoing a stress test. The \WMAP\ measurements of the  primary CMB first established that the $\Lambda$ Cold Dark Matter (\LCDM) model with just six parameters is an excellent fit to CMB measurements of the radiation anisotropies of the universe~\citep{2003ApJS..148..175S,2013ApJS..208...19H}. Measurements from \Planck\ have reinforced this model \citep{1807.06205}.  Distinct probes of the geometry, expansion, and growth of structure from a wide range of cosmic epochs have now reached percent-level precision. Many are consistent with the \LCDM\ model derived from the primary CMB anisotropy in the early universe \citep[e.g.,][]{1907.05922,2007.08991,hsc,2203.07128,2206.08327,2211.16794,2303.05537} but some are in tension, with varying levels of significance. A local measurement of the expansion rate, calibrated using Cepheid variable stars, is 7\% higher than the prediction from \Planck\ assuming the \LCDM\ model \citep{2112.04510}, at quoted $5\sigma$ significance. Many measurements of structure growth are $\simeq 10$\% lower than what the standard model based on \Planck\ parameters predicts \citep{1611.08606,hikage19,2007.15632,2010.00466,2105.12108,2105.13549,2110.10141,2111.09898,2203.12440}, at 2–$3\sigma$ significance. At the same time, increasingly precise measurements of late-universe observables are quickly opening up a path towards constraining extensions of the standard model, including the mass of neutrinos and the equation of state for the dark energy component purported to cause cosmic acceleration. Ground-based CMB surveys like ACT and SPT, with their high angular resolution, are uniquely positioned to weigh in on these issues from multiple fronts, expanding on the \Planck\ legacy.

In this work, we use ACT Data Release 6 (DR6) to measure gravitational lensing of the CMB and produce a mass map covering $9400\,\si{deg}^2$. We combine the power spectrum of the fluctuations in this map with measurements of the baryon acoustic oscillations (BAO) measured by 6dF \citep{1106.3366} and the Sloan Digital Sky Survey \citep[SDSS;][]{10.1086/342343,10.1088/0004-6256/142/3/72,10.1088/0004-6256/145/1/10} to obtain one of the most precise measurements to date of the amplitude of matter fluctuations. Our combination of ACT and \Planck~ lensing along with BAO, in particular, provides a state-of-the art view of structure formation.  The first question we ask is whether the amplitude of matter fluctuations is lower than the early-universe prediction from \Planck\  and whether it is in agreement with other late-time measurements (such as optical weak lensing), which probe lower redshifts than CMB lensing does. Here, we use our new CMB lensing data to measure the mass fluctuations, primarily from linear scales, dominated by the structures at redshifts $z=0.5$--$5$. We also present a suite of constraints on several extensions to the standard cosmological model including the sum of masses of neutrinos and deviations of the spatial curvature of the universe from flatness.

This paper is one out of a larger set of papers on ACT DR6. It presents our CMB lensing mass map and explores the consequences for cosmology from the combination and comparison of our lensing measurements with other external data (including those in the context of extensions to the \LCDM\ model). In \cite{Qu23}, we present the measurement of the CMB lensing power spectrum used in the cosmological constraints of this work, with details on the data analysis and verification pipeline. \cite{Qu23} also presents constraints on cosmological parameters from ACT CMB lensing alone, such as $\scmbl\equiv\seightc$. \cite{dr6-lensing-fgs} provides a detailed investigation of the characterization and mitigation of our most significant systematic in the lensing power spectrum measurement: the bias due to extragalactic astrophysical foregrounds. 

\section{A wide-area high-fidelity mass map}\label{sec:map}
  
\begin{figure}
\includegraphics[width=\columnwidth]{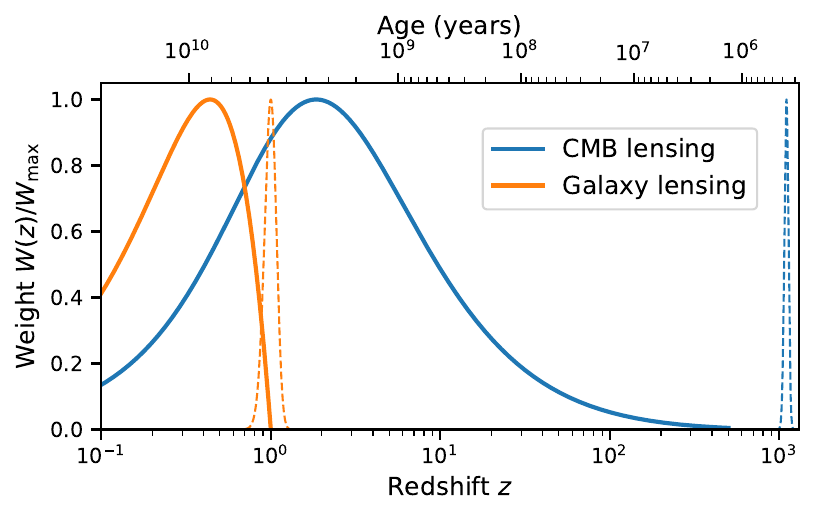}
  \caption{Mass-map weights for CMB and galaxy weak lensing, normalized to the maximum value. The blue solid curve shows the relative weights different redshifts receive in a mass map reconstructed from CMB lensing (as in this work) and the orange solid curve shows the same for a sample of galaxies at $z=1$ (typical of current galaxy lensing surveys). The dashed curves show the corresponding source distribution, with that for the CMB centered at the redshift of last-scattering around $z=1100$. The comoving distances to the peak redshifts are roughly 1\,Gpc (galaxy lensing) and 5\,Gpc (CMB lensing). An angular scale of $\sim 1 \deg$ or a lens multipole of $L=200$ then corresponds to comoving wave-numbers at those distances of roughly 0.2\,Mpc$^{-1}$ (galaxy lensing) and 0.04\,Mpc$^{-1}$ (CMB lensing). }
    \label{fig:weight}
   
\end{figure}

High-resolution measurements of the CMB allow us to reconstruct a map of CMB gravitational lensing convergence; this provides a view of the mass distorting the CMB (emitted from the last-scattering surface) due to its gravitational influence. The convergence directly probes the total mass density of the universe integrated along the line-of-sight all the way to the redshift of recombination $z_\star \simeq 1100$, although nearly all of the contribution comes from redshifts $z<30$, with peak contributions around $z=0.5$--$5$. The convergence is related to the underlying total matter overdensity $\delta_{\rm m}(\boldsymbol{x}) = (\rho_{\rm m}(\boldsymbol{x})-\bar{\rho}_{\rm m})/\bar{\rho}_{\rm m}$ (where $\rho_{\rm m}(\boldsymbol{x})$ is the matter density and $\bar{\rho}_{\rm m}$ is the mean matter density) through
\begin{equation}
    \kappa(\mathrm{\bf{\hat{n}}}) = \int_0^{\infty} dz W^\kappa(z) \\
                \delta_{\rm m}(\chi(z)\mathrm{\bf{\hat{n}}}, z).
\end{equation}
In the case of a flat universe with zero spatial curvature, the lensing kernel $W^\kappa$ simplifies to

\begin{equation}
    W^\kappa(z) = \frac{3}{2}\Omega_{m} H_0^2 \\
            \frac{(1+z)}{H(z)} \frac{\chi(z)}{c} \int_z^{\infty} dz' n(z') \\
                \frac{\chi(z') - \chi(z)}{\chi(z')},
\end{equation}
where $n(z)$ is the normalized redshift distribution of the light source undergoing gravitational lensing and $\chi(z)$ is the comoving distance to redshift $z$. While this expression is general (e.g., as appears in cosmic shear distortions of galaxy shapes, see \citealp{1710.03235}), when the lensed light source is the CMB, the redshift distribution can be approximated as $n(z) \simeq \delta^D(z-z_\star)$, where $z_\star \simeq
1100$ is the redshift of the surface of last scattering and $\delta^D$ is the
Dirac delta function. Thus, for the CMB lensing mass maps produced here, we have \citep{0601594}

\begin{equation}
    W^{\kcmb}(z) =  \frac{3}{2}\Omega_{m}H_0^2  \frac{(1+z)}{H(z)} \frac{\chi(z)}{c} \\ 
    \left [ \frac{\chi(z_\star)-\chi(z)}{\chi(z_\star)} \right ].
    \label{eqn:cmb_kernel}
\end{equation}

In Figure \ref{fig:weight}, we compare the lensing weight kernels for CMB lensing and an illustrative sample of galaxies at $z=1$, a typical source redshift for current galaxy lensing surveys. CMB lensing provides a complementary view of epochs of the late-time universe that are otherwise difficult to access with galaxy surveys while also significantly overlapping with low-redshift surveys, allowing for a rich variety of cross-correlation analyses. 

{\bf ACT DR6 data:} The mass map and cosmological parameters in this work are derived from CMB data from ACT. Located on Cerro Toco in the Atacama Desert in northern Chile, ACT observed the sky at millimeter wavelengths from 2007 until 2022. From 2016, the telescope was equipped with the Advanced ACTPol (AdvACT) receiver containing arrays of superconducting transition-edge sensor bolometers, sensitive to both temperature and polarization at frequencies centered roughly at 30, 40, 97, 149 and $225\,\si{GHz}$ \citep{fowler2007optical,thornton2016atacama}; we denote these bands as \texttt{f030}, \texttt{f040}, \texttt{f090}, \texttt{f150} and \texttt{f220}. Our current analysis uses night-time temperature and polarization AdvACT data collected from 2017 to 2021 covering the CMB-dominated frequency bands \texttt{f090} and \texttt{f150}, constituting roughly half of the total volume of data collected by ACT since its inception.  Here, we use an early science-grade version of the ACT DR6 maps, labeled \texttt{dr6.01}. Since the maps used in our analysis were generated, we have made some refinements to the map-making that improve the large-scale transfer function and polarization noise levels, and include data taken in 2022, although we have performed extensive testing in \citealt{Qu23} to ensure that the \texttt{dr6.01} map quality is sufficient for lensing analysis. We anticipate using a future version of these maps for further science analyses and for the DR6 public data release. Additionally, data collected during the daytime, at other frequency bands, and during the years 2007--2016 are also not included in the lensing measurement presented here, but we intend to include them in a future analysis.

{\bf Software and pipeline:} In order to transform maps of the CMB to maps of the lensing convergence, a preliminary publicly available and open-source pipeline has been developed for the upcoming Simons Observatory \citep[SO;][]{1808.07445}; we demonstrate this pipeline for the first time on ACT data in this series of papers. The SO stack consists of the pipeline code \href{https://github.com/simonsobs/so-lenspipe/}{\texttt{so-lenspipe}}, which depends primarily on a reconstruction code \href{https://github.com/simonsobs/falafel/}{\texttt{falafel}}, a normalization code \href{https://github.com/simonsobs/tempura/}{\texttt{tempura}}, and the map manipulation library \href{https://github.com/simonsobs/pixell/}{\texttt{pixell}}. We briefly summarize the measurement here, but the details can be found in our companion paper,~\citet{Qu23}.

\begin{figure}
\includegraphics[width=\columnwidth]{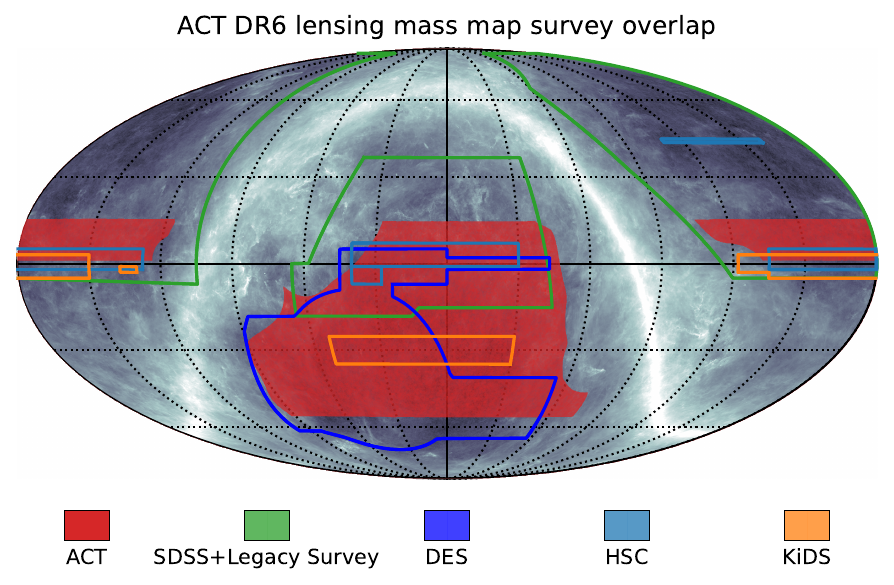}
  \caption{Overlap of the ACT mass map (red) with various ongoing galaxy surveys. The green contours show a rough union of the footprint of SDSS, the DECam Legacy Survey and the Mayall z-band Legacy Survey, with Dark Energy Spectroscopic Instrument (DESI) data expected to be available in part of this region \citep{1807.09287,10.3847/1538-3881/ab089d}. The grayscale background is a Galactic dust map from \Planck\ \citep{1502.01588}.  }
    \label{fig:foot}
\end{figure}

\begin{figure*}
\includegraphics[trim={0 2cm 0 1.2cm},clip,width=\textwidth]{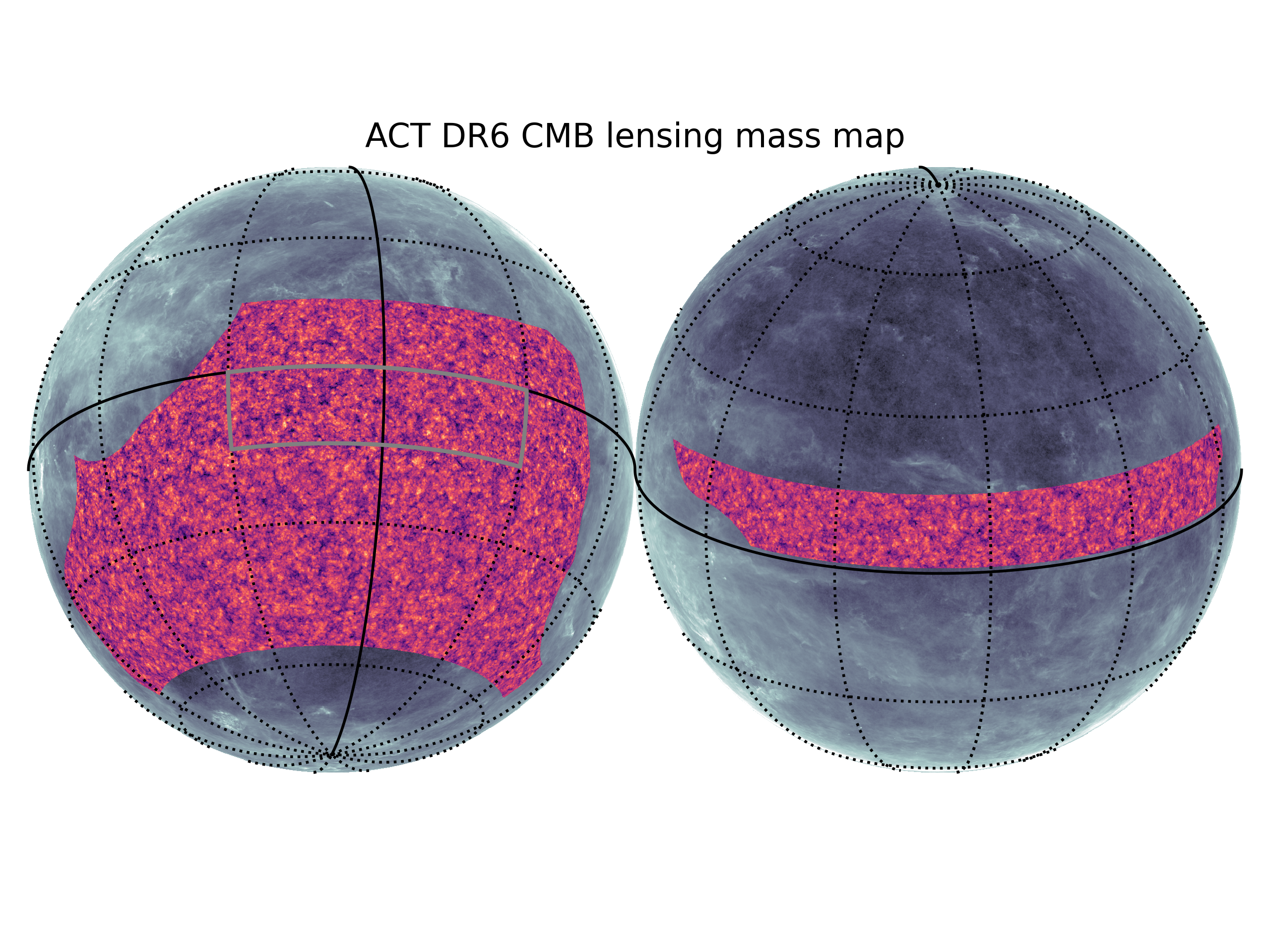}
  \caption{ACT DR6 CMB lensing mass map presented in this work. The map covers $9400\,\si{deg}^2$ or sky fraction $f_{\rm sky}=0.23$ with a signal-to-noise significantly greater than unity over a wide range of scales. We show the Wiener-filtered CMB lensing convergence in an orthographic projection with bright orange corresponding to peaks of the dark-matter dominated mass distribution and dark purple regions corresponding to voids. Dark-matter dominated structures on few-degree scales corresponding to the peak of the lensing power spectrum can be seen by eye (see also Figure \ref{fig:Nell}).  The grayscale background is a Galactic dust map from \Planck\ \citep{1502.01588}; our analysis mask is designed to avoid dusty regions of the sky. The region in the gray box is shown in Figure \ref{fig:zoom}.}
    \label{fig:lmap}
\end{figure*}

\begin{figure*}
\includegraphics[width=\textwidth]{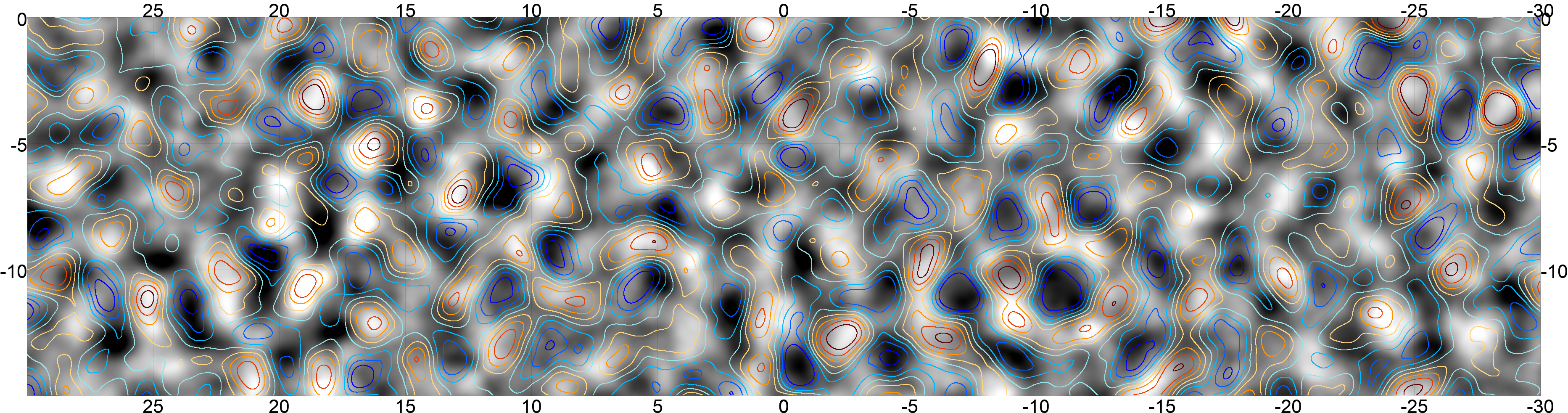}
  \caption{A zoom-in of a $900\,\si{deg}^2$ region of the ACT DR6 mass map shown as the Wiener-filtered gravitational potential (related to the convergence through $\nabla^2\phi=-2\kappa$). The distribution of dusty galaxies constituting the CIB measured by \Planck~ is overlaid as contours. The overdensities in red correspond well with the bright/white mass-dominated regions of the mass map and the underdensities in blue correspond well with the darker mass-devoid regions.}
    \label{fig:zoom}
\end{figure*}

{\bf Producing a lensing map:} The individual frequency maps are pre-processed and inverse-variance co-added. At \texttt{f090} and \texttt{f150}, the maps have an average white noise level of 16 and $17\,\mu\text{K-arcmin}$, respectively, though there is considerable contribution from correlated atmospheric noise on the largest scales (around $0.3 \deg$) used in our analysis as well as moderate levels of inhomogeneity (see \citealp{2111.01319} and \citealp{2303.04180} for details of ACT noise).  We use the quadratic estimator formalism \citep{Okamoto2003,1807.06210} to transform maps of the co-added CMB (whose harmonic transform modes we represent with $\ell$) to maps of the lensing convergence (whose harmonic transform modes we represent with $L$); this formalism exploits the fact that gravitational lensing couples previously independent spherical harmonic modes of the unlensed CMB in a well-understood way. We exclude scales in the input CMB maps with multipoles $\ell<600$ since these contain significant atmospheric noise and Galactic foregrounds. We exclude small scales (multipoles $\ell>3000$) due to possible contamination from astrophysical foregrounds like the thermal Sunyaev--Zeldovich (tSZ) effect, the cosmic infrared background (CIB), the kinetic SZ (kSZ) effect, and radio sources.  Crucially, we perform ``profile hardening'' on this estimator \citep{2007.04325}, a variation of the ``bias hardening'' procedure \citep{1209.0091,Osborne}. This involves constructing a quadratic estimator reconstruction designed to capture mode-couplings arising from objects with radial profiles similar to the tSZ imprints of galaxy clusters. We then construct a linear combination of the usual lensing estimator with this profile estimator such that the response to the latter is nulled. The deprojection of contaminants using this profile hardening approach is our baseline method for mitigation of  contamination from extragalactic astrophysical foregrounds, though we also obtain consistent results with alternative mitigation schemes, e.g. involving spectral deprojection of foregrounds \citep{1802.08230,2004.01139} and shear estimation \citep{1804.06403,2208.14988}. The companion paper \citet{dr6-lensing-fgs} investigates in detail the bias from foregrounds and shows how our baseline choice fully mitigates the bias from all known sources of foregrounds (including the CIB).

Additionally, our mass maps are made using a novel cross-correlation-based estimator \citep{Madhavacheril2021}: this is a modification of the standard quadratic estimator procedure \citep{Okamoto2003} that, through the use of time-interleaved splits, only includes terms that have independent instrument noise.  This makes our measurement insensitive to mismodeling of instrument noise.\footnote{This is optimized for current and forthcoming ground-based surveys, which have complicated noise properties due to the interplay between the atmospheric noise and the telescope scanning strategy.} For the released mass map in particular, this ensures that a `mean-field' term we subtract to correct for mask- and noise-induced statistical anisotropy (see e.g, \citealp{10.1051/0004-6361/201321048}) does not depend on details of the ACT instrument noise, allowing for the scatter in cross-correlations on large angular scales to be predicted more reliably.

While scales with multipoles $600 < \ell < 3000$ are used from the input CMB maps, the output lensing mass maps are made available on larger scales, down to lower multipoles $L$; this is possible due to the way large-scale lenses coherently induce distortions in the small-scale CMB fluctuations.  For the same reason, most of the information in the lensing reconstruction process comes from small angular scales in the CMB maps with multipoles $\ell>1500$.

\begin{figure*}
    \centering
    \includegraphics[width=\textwidth]{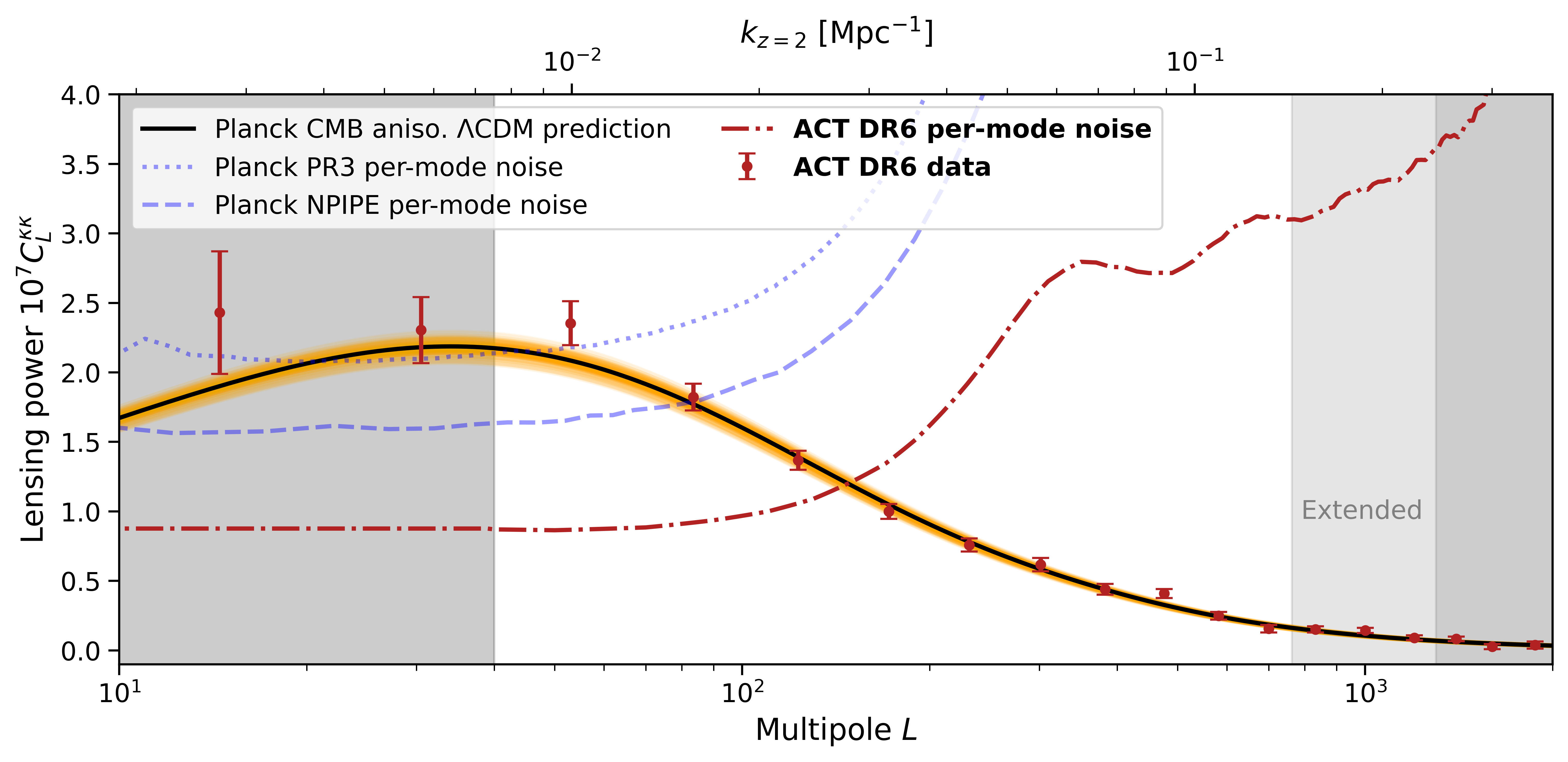}
    \caption{The ACT DR6 CMB lensing power spectrum measurement, from \cite{Qu23}. The bandpowers of the two-point statistics of the DR6 mass map are shown as red data points. The black solid curve shows the {\it prediction} for this signal in the $\Lambda$CDM model based on the measurement of the primary CMB anisotropies by the \Planck\ satellite; i.e., this prediction is not a fit to the ACT data. The prediction and our measurement (presented in detail in the companion paper, \citealp{Qu23}) are in excellent agreement, showing the success of the \LCDM\ model in propagating a measurement of the radiation anisotropies at age of the universe $t\simeq 375,000$ years ($z\simeq 1100$) to the matter fluctuations at $t\simeq 1-9$ billion years ($z\simeq 0.5-5$). We also show samples (orange) from $\Lambda$CDM chains of the \Planck~ primary CMB anisotropy measurements to highlight the uncertainty in the early universe prediction. The dotted, dashed, and dot-dashed curves show the noise power spectra (i.e., the variance of the reconstruction noise per mode) in the mass maps produced by \Planck\ PR3 \citep{1807.06210}, \Planck\ \NPIPE\ \citep{2206.07773}, and this work, respectively. The ACT mass map is signal-dominated out to $L\simeq 150$, providing a high-fidelity view of the dark-matter-dominated mass distribution. The dark gray regions are not included in our analysis and the light gray region is included in our ``extended'' analyses. The top axis shows the comoving wave-number $k=L/\chi(z_p)$ at the peak redshift of the CMB lensing kernel $z_p=2$. }
    \label{fig:Nell}
\end{figure*}

{\bf Mass-map properties:} Covering a fraction $f_{\rm sky}\simeq 0.23$ of the full sky, the ACT DR6 CMB lensing mass map overlaps with a number of large-scale structure surveys, providing opportunities for cross-correlations and joint analyses (see Figure \ref{fig:foot}). In Figure \ref{fig:lmap}, we show a visual representation of the mass map in an orthographic projection with bright orange corresponding to peaks in the dark matter-dominated mass distribution and dark purple corresponding to voids in the mass distribution. We also show in Figure \ref{fig:zoom} a zoom-in of a $900\,\si{deg}^2$ region of the mass map in grayscale (bright regions being peaks in the mass and dark regions being voids) overlaid with a map of the CIB constructed by the \Planck~ collaboration using measurements of the millimeter sky at $545\,\si{GHz}$ \citep{1502.01588}. The CIB consists primarily of dusty star-forming galaxies with contributions to the emissivity peaking around $z=2$ when star formation was highly efficient. Since this also happens to be where the CMB lensing kernel peaks, the CMB lensing maps and the CIB are highly correlated. The high correlation coefficient and the high per-mode signal-to-noise ratio of the ACT mass maps allows us to see by eye the correspondence of the dark-matter dominated mass reconstruction in grayscale and the CIB density in colored contours.

In Figure \ref{fig:Nell}, we show the power spectra of the reconstruction noise for various mass maps from \Planck~(which cover 65\% of the sky) against the noise power spectrum of the ACT DR6 mass map. The ACT map is signal-dominated on scales $L<150$, similar to the \texttt{D56} maps from the ACT DR4 release \citep{2004.01139}, but covering 20 times more area.  In comparison to the \Planck~ maps, the ACT mass map has a reconstruction noise power that is at least a factor of two lower, although we note that the \Planck\ maps cover more than twice the area. The small scales are reconstructed with much better precision than \Planck, allowing the ACT mass map to be of particular use in the `halo lensing' 
 regime for cross-correlations with galaxy groups \citep[e.g.,][]{1411.7999,10.1103/PhysRevD.98.043506} and galaxy clusters \citep[e.g.,][]{1412.7521,1502.01597,1708.01360,10.1038/s41550-017-0259-1,1907.08605,2009.07772}. There are some associated caveats in this regime that we describe in Section \ref{sec:clusters}. Our mass map is also highly complementary to that from galaxy weak lensing with DES \cite{1708.01535,2105.13539}, which uses source galaxies at redshifts up to $z\simeq 1.5$. This map covers around $4100\,\si{deg}^2$ and has significant overlap with the DR6 ACT CMB lensing mass map (see Figure \ref{fig:foot}).

When using the ACT mass map in cross-correlation, we do not recommend using scales with multipoles $L<40$, since the null and consistency tests from \cite{Qu23} suggest that those scales may not be reliable. Similarly, we find evidence in \cite{dr6-lensing-fgs} that multipoles $L>1300$ may not be reliable from the perspective of astrophysical foreground contamination. However, the precise maximum multipole to be used in cross-correlations will be dictated both by theory modeling concerns as well as improved assessments of foreground contamination specific to the cross-correlation of interest. We enable investigations of the latter by providing a suite of simulated reconstructions that include foregrounds from the Websky extragalactic foreground simulations~\citep{2001.08787}.

{\bf Lensing power spectrum: } To obtain cosmological information from the mass map, we compute its power spectrum or two-point function. Since the mass map is constructed through a quadratic estimator, and hence, has two powers of the CMB maps, the power spectrum is effectively a four-point measurement in the CMB map. This four-point measurement requires subtraction of a number of biases in order to isolate the component due to gravitational lensing. The largest of these biases is the Gaussian disconnected bias, which depends on the two-point power spectrum of the observed CMB maps and is thus non-zero even in the absence of lensing. As discussed in detail in \cite{Qu23}, the use of a cross-correlation-based estimator \citep{Madhavacheril2021} adds significantly to the robustness of our measurement since the large Gaussian disconnected bias we subtract (see, e.g., \citealp{1008.4403}) using simulations does not depend on the details of ACT instrumental noise. This novel estimator also significantly reduces the computational burden in performing null tests (which have no Gaussian disconnected bias from the CMB signal in the standard estimator), since the expensive simulation-based Gaussian bias subtraction can be skipped altogether.

The CMB lensing power spectrum from \cite{Qu23} is determined at $2.3\%$ precision, corresponding to a measurement signal-to-noise ratio of $43\sigma$. To our knowledge, this measurement is competitive with any other weak lensing measurement, with precision comparable to that from \Planck\ \citep{2206.07773} and with complementary information on smaller scales $L>400$. In \cite{Qu23}, we verify our measurements with an extensive suite of $\mathcal{O}(100)$ map-level and power-spectrum-level null tests and find no evidence of systematic biases in our measurement. These tests include splitting the data by multipole ranges, detector array, frequency band, and inclusion of polarization, as well as variation of regions of the sky masked. 

Our analysis followed a blinding policy where no comparisons with previous measurements or theory predictions were allowed until the null tests were passed. Unless otherwise mentioned, the results in this work are based on the `baseline' multipole range of $40<L<763$ decided before unblinding. In some cases, we also provide runs with an `extended' multipole range of $40 < L < 1300$, which was deemed to be reliable following a re-assessment of foreground biases from simulations \citep{dr6-lensing-fgs} that was done post-unblinding. 

\section{Is the amplitude of matter fluctuations  low?}\label{sec:s8}

We next use the power spectrum of the mass map to characterize the amplitude of matter fluctuations. This allows us to compare our measurement with those from other cosmological probes of structure formation such as galaxy cosmic shear. We focus on the parameter $\sigma_8$, which is formally the root-mean-square fluctuation in the {\it linear} matter overdensity smoothed on scales of $8\,\si{Mpc}/h$ at the present time.\footnote{In our companion paper, \cite{Qu23}, we fit for the parameter combination $\scmbl=\seightc$; here we isolate $\sigma_8$ in order to compare with galaxy weak lensing, which has a different scaling with the matter density $\Omega_{\rm m}$.} Fitting for this parameter therefore requires propagating a model prediction for the linear growth of matter fluctuations over cosmic time to the observed matter power spectrum (projected along the line-of-sight when using lensing observables).

Different probes of the late universe access different redshifts or cosmic epochs (see Figure \ref{fig:weight}) and are also sensitive to different scales. Consequently,  differences among the inferred values of $\sigma_8$ from various late-universe probes or with the early-universe prediction based on CMB anisotropies can hint at possibilities such as: (a) non-standard redshift evolution of the growth of structure, possibly due to modifications of general relativity (e.g.,~\citealp{2107.12992,2302.01331}); (b) a non-standard power spectrum of matter fluctuations, e.g., due to axion dark matter (e.g.,~\citealp{2301.08361}) or dark matter-baryon scattering (e.g.,~\citealp{2301.08260}); (c) incorrect modeling of small-scale fluctuations, e.g., due to non-linear biasing (for galaxy observables) or baryonic feedback (for lensing observables; e.g.,~\citealp{2206.11794}); or (d) unaccounted systematic effects in one or more of these measurements. By providing a measurement of $\sigma_8$ with CMB lensing, we probe mainly linear scales with information from a broad range of redshifts $z\sim0.5$--$5$, which peaks around $z=2$ as shown in Figure \ref{fig:weight}. 

\begin{table}
\caption{Parameters and priors used in this work. See Section~\ref{sec:s8} for definitions of the parameters. Uniform priors are shown in square brackets and Gaussian priors with mean $\mu$ and standard deviation $\sigma$ are shown as $\mathcal{N}(\mu,\sigma)$. In all cases, we additionally reject samples where the derived parameter $H_0$ falls outside the range $[\hmin, \hmax]\, \text{km}\,\text{s}^{-1}\,\text{Mpc}^{-1}$. These prior choices closely follow those in \Planck\ analyses \citep{1502.01591,1807.06210,2206.07773}. }
\label{tab:params}
\begin{tabular*}{\columnwidth}{ l  @{\extracolsep{\fill}} c  c}
\hline
\hline
Parameter   & Prior  \\ \hline
\multicolumn{2}{l}{\bf Lensing + BAO } \\
$\omch$  & $[\omchmin, \omchmax]$ \\ 
$\ombh$    & $\mathcal{N}(\obmean,\obsigma)$  \\ 
$\loga$   & $[\logamin,\logamax]$  \\ 
$\ns$ & $\mathcal{N}(\nsmean,\nssigma)$  \\
$\thetamc$  & $[\thetamin, \thetamax]$   \\
\hline
\multicolumn{2}{l}{\bf Lensing + BAO + CMB anisotropies } \\
$\omch$  & $[\omchmin, \omchmax]$  \\ 
$\ombh$    & $[\ombhmin, \ombhmax]$  \\ 
$\loga$   & $[\logamin,\logamax]$  \\ 
$\ns$ & $[\nsmin, \nsmax]$  \\
$\thetamc$  & $[\thetamin, \thetamax]$   \\
$\tau$    & $[\taumin, \taumax]$ \\
\hline
\multicolumn{2}{l}{\bf Lensing + BAO + CMB anisotropies. $\Lambda$CDM}\\
\multicolumn{2}{l}{\bf extensions include the above six and one of below} \\
$\mnu$ (eV)   & $[\mnumin, \mnumax]$   \\
$\omegak$  & $[\okmin, \okmax]$   \\
\hline
\end{tabular*}

\end{table}

We set up a likelihood and inference framework for cosmological parameters detailed in Appendix~\ref{app:like}, considering a spatially flat \LCDM\ universe and freeing up the five cosmological parameters shown in the first section of Table \ref{tab:params}: the physical cold dark matter density, $\omch$; the physical baryon density, $\ombh$; the amplitude of scalar primordial fluctuations, $\loga$; the spectral index of scalar primordial fluctuations, $\ns$; and the approximation to the angular scale of the sound horizon at recombination used in \href{https://cosmologist.info/cosmomc/}{\texttt{CosmoMC}}, $\thetamc$. We note that we have an informative prior on $n_{\text{s}}$, which is necessary, since the spectral index and the amplitude of fluctuations are degenerate given only a measurement of the lensing power spectrum. As noted in \Planck\ CMB lensing analyses \citep{1502.01591,1807.06210}, constraints on the amplitude of fluctuations are only weakly sensitive to the choice of this prior within plausible bounds informed by CMB anisotropies.   In particular, this prior is centered on but also five times broader than the constraint obtained from \Planck\ measurements of the CMB anisotropy power spectra in the \LCDM\ model~\citep{1807.06209}, and two times broader than constraints obtained there from various extensions of \LCDM. This prior is, therefore, quite conservative.  The prior on the baryon density $\ombh$ we use is from updated Big Bang Nucleosynthesis (BBN) measurements of deuterium abundance from \cite{10.1038/s41586-020-2878-4}, but the constraints are not noticeably degraded using broader priors, e.g., from \cite{1710.11129}.

Importantly, in our comparison here of CMB lensing, galaxy weak lensing, and CMB anisotropies, we fix the sum of neutrino masses $\mnu$ to be the minimal value of $0.06\,\si{eV}$ allowed by neutrino oscillation experiments (with one massive and two massless neutrinos), but we return to constraining this parameter with ACT data in Section~\ref{sec:mnu}. We also compare our results from CMB lensing with those from the two-point power spectrum of the CMB anisotropies themselves; see Appendix~\ref{app:cmb} for details on constraints from the latter that we revisit with our inference framework.

\subsection{BAO likelihoods}
\label{sec:bao}

Weak lensing measurements depend primarily on the amplitude of matter fluctuations $\sigma_8$, the matter density $\Omega_{\rm m}$, and the Hubble constant $H_0$. In order to reduce degeneracies of our $\sigma_8$ constraint with the latter parameters and allow for more powerful comparisons of lensing probes with different degeneracy directions, we include information from the 6dF and SDSS surveys. The data we include measures the BAO signature in the clustering of galaxies with samples spanning redshifts up to $z\simeq 1$, including 6dFGS \citep{1106.3366}, SDSS DR7 Main Galaxy Sample (MGS;~\citealt{1409.3242}), BOSS DR12 luminous red galaxies (LRGs; \citealp{1607.03155}), and eBOSS DR16 LRGs \citep{2007.08991}. We do not use the higher-redshift Emission Line Galaxy \citep[ELG; ][]{1509.05045}, Lyman-$\alpha$ \citep{2007.08995}, and quasar samples \citep{2007.08998}, though we hope to include these in future analyses. We only include the BAO information from these surveys (which provides constraints in the $\Omega_{\rm m}$--$H_0$ plane) and do not include the structure growth information in the redshift-space distortion (RSD) component of galaxy clustering. We make this choice so as to isolate information on structure formation purely from lensing alone.

\subsection{The ACT lensing measurement of $\sigma_8$}

\begin{figure*}
    \centering
    \includegraphics[width=0.49\textwidth]{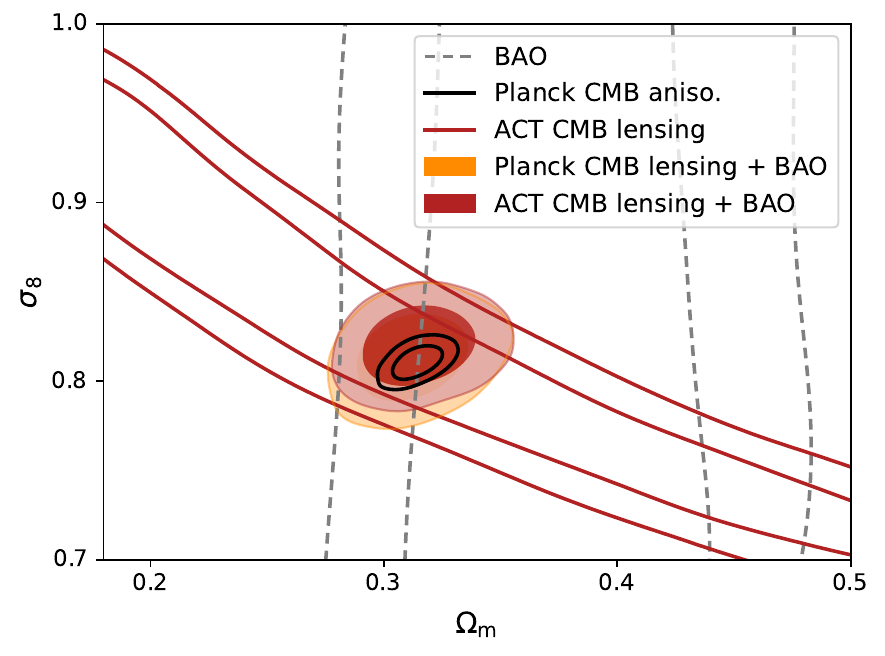}
    \includegraphics[width=0.49\textwidth]{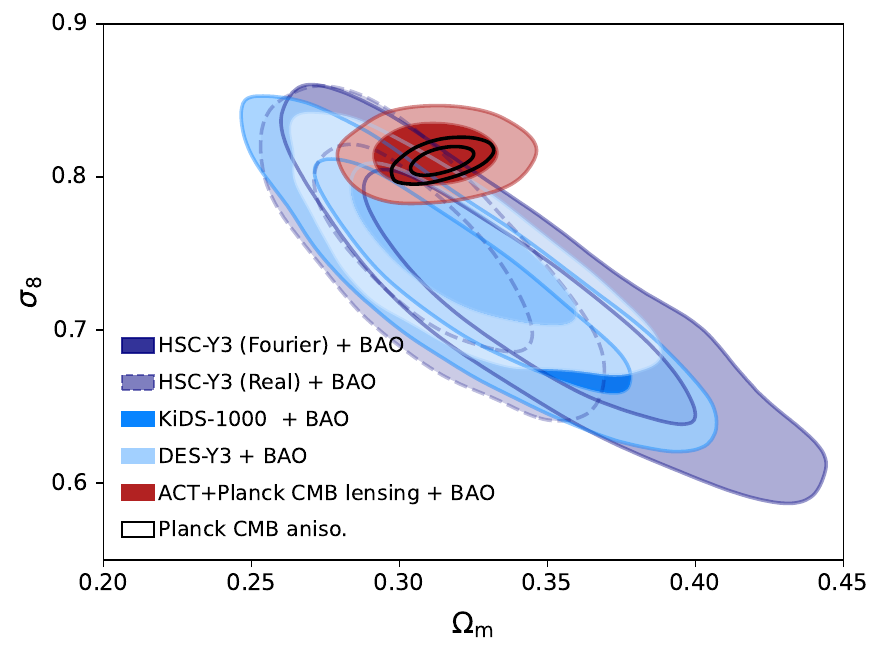}
    \caption{{\it (a) Left}: The ACT lensing measurement of the amplitude of matter fluctuations $\sigma_8$. For each data set, we show 68\% and 95\% confidence limits. Lensing measurements also depend on $H_0$ and $\omegam$; we break this degeneracy by including BAO data. The ACT lensing measurement agrees well with the \Planck\ lensing measurement as well as the inference of $\sigma_8$ from \Planck\ CMB anisotropies assuming \LCDM, a mainly early-universe measurement. {\it (b) Right}: Comparison of $\sigma_8$ measurements between ACT CMB lensing and a consistent re-analysis of galaxy weak lensing (cosmic shear) data sets. The latter also are degenerate with other parameters (more severely; see Appendix~\ref{app:tubes}). All constraints here -- except those from \Planck\ CMB anisotropies -- include a BBN prior on $\Omega_{\rm b} h^2$.}
    \label{fig:sig8om}
\end{figure*}

\begin{figure*}
    \centering
    \includegraphics[width=0.99\textwidth]{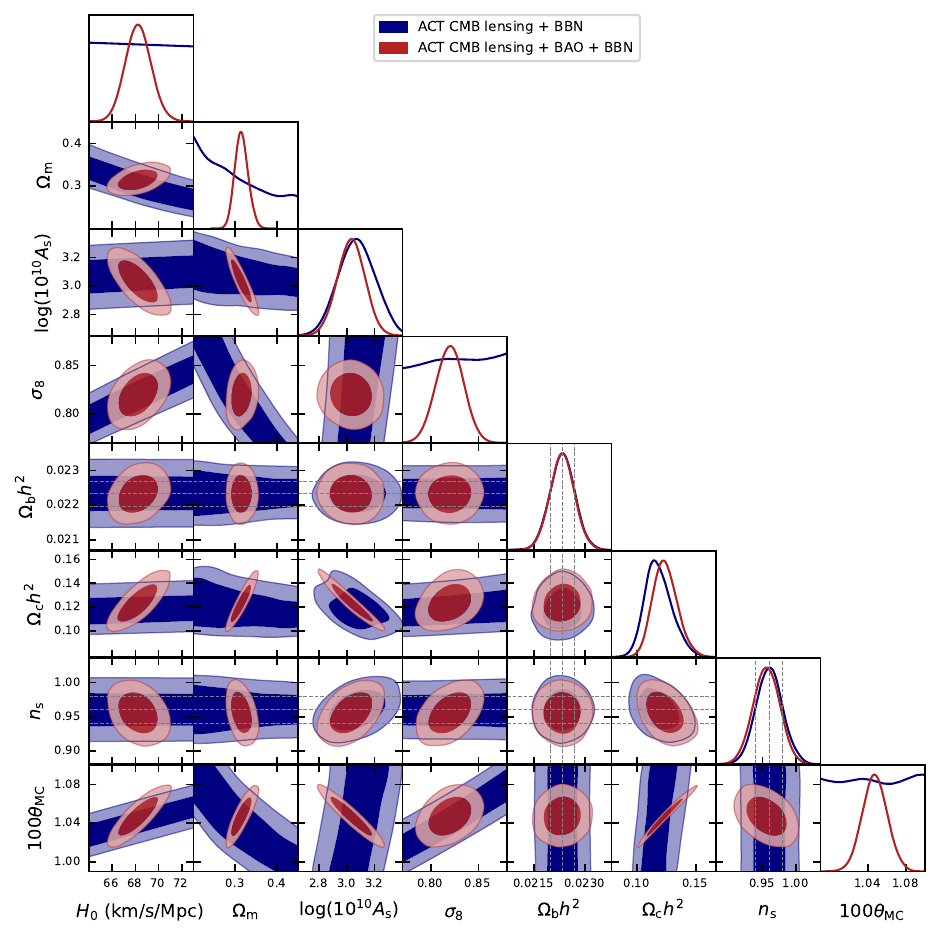}
    \caption{Marginalized 2d and 1d posteriors for ACT CMB lensing either in combination with a BBN prior on $\Omega_{\rm b}h^2$ (blue; as done in \cite{Qu23}) or in combination with both BBN and galaxy BAO (red). The parameters $H_0$, $\Omega_m$ and $\sigma_8$ are derived, while the remaining are sampled. Informative priors on $\Omega_{\rm b}h^2$ and $n_s$ are indicated as vertical lines (68\% c.l. and mean of priors); all other priors lie well outside the plotted region.  }
    \label{fig:tri}
\end{figure*}

\begin{figure}
    \centering
    \includegraphics[width=\columnwidth]{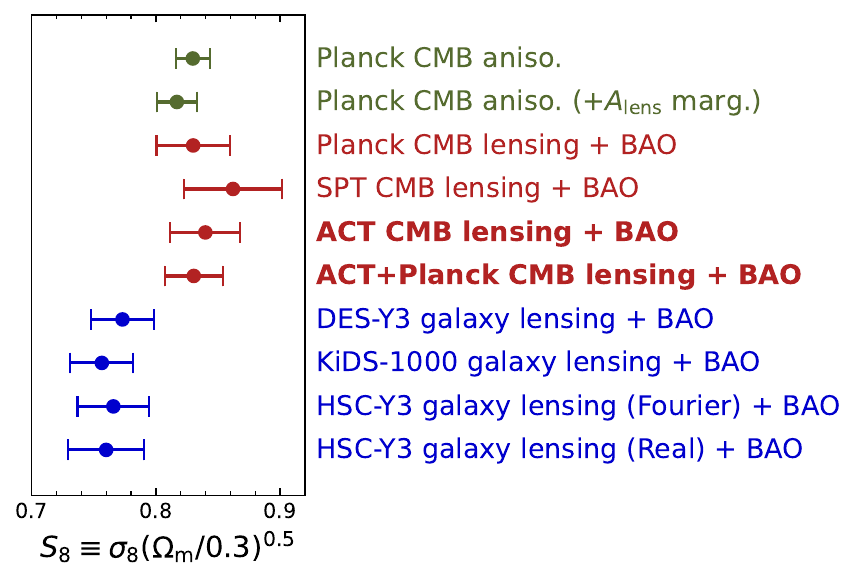}
    \includegraphics[width=\columnwidth]{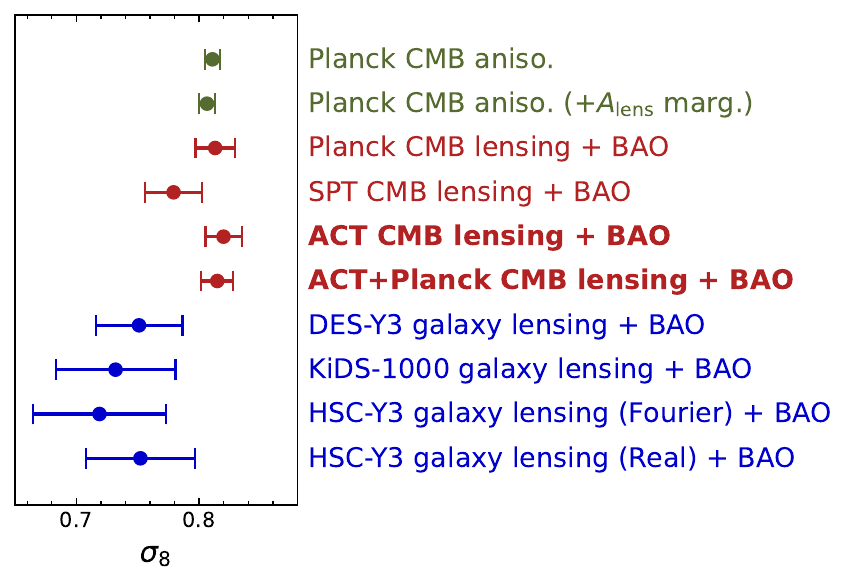}
    \caption{Marginalized posteriors for various combinations of parameters measuring the amplitude of matter fluctuations. The top panel shows $S_8\equiv \seightg$ which is best constrained by galaxy lensing, and the bottom panel shows $\sigma_8$. All lensing measurements shown here include BAO data. The \Planck\ CMB anisotropy measurements are shown both without and with marginalization over late-time information; while the former is mostly an early-universe extrapolation, the latter is more fully so. }
    \label{fig:S8_bao}
\end{figure}

The ACT lensing power spectrum shown in Figure \ref{fig:Nell} is  proportional on large scales to the square of the amplitude of matter fluctuations $\sigma_8$ and is therefore an excellent probe of structure growth. This is particularly so in combination with BAO, which does not measure structure growth but whose expansion history information helps break degeneracies with $\omegam$ and $H_0$. In Figure \ref{fig:sig8om}, we show constraints in the $\sigma_8$--$\omegam$ plane, and in Figure \ref{fig:tri} we show all the sampled parameters. The gray dashed contours from BAO alone do not provide information in the $\sigma_8$ direction and the ACT lensing-alone data-set constrains well roughly the parameter combination $\seightc$ \citep[see][for further investigation of this combination]{Qu23}. The combination of ACT lensing and BAO provides the following \aseightprecision\% marginalized constraint (see Table \ref{tab:marg}):
\begin{equation}
\sigma_8= \aseightmean \pm{\aseighterr} .
\end{equation}
This is consistent with the value inferred from \Planck\ measurements of the CMB anisotropies that mainly probe the early universe, as can also be seen in the marginalized constraints in Figure \ref{fig:S8_bao}.  Since CMB anisotropy power spectra also contain some information on the late-time universe (primarily through the smoothing of the acoustic peaks due to lensing), we additionally show inferred values of $\sigma_8$ where the lensing information has been marginalized over (by freeing the parameter $A_{\rm lens}$; \citealp{0803.2309})\footnote{In this paper, as in \cite{0803.2309}, we use $A_{\rm lens}$ to refer to an amplitude scaling of the lensing that induces smearing of acoustic peaks in the 2-point power spectrum while leaving the 4-point lensing power spectrum fixed. We caution that the same notation is used in \cite{Qu23} for a different parameter characterizing the amplitude of the measured 4-point lensing power spectrum with respect to a prediction using a \LCDM\ cosmology that best fits the \Planck\ CMB anisotropies.  } so as to isolate the early-universe prediction from \Planck\ (see Appendix~\ref{app:cmb} for more information). Our CMB-lensing-inferred late-time measurement remains consistent with this $A_{\rm lens}$-marginalized prediction of $\sigma_8$ from the \Planck\ CMB anisotropies.

\begin{figure}[t]
    \centering
    \includegraphics[width=\columnwidth]{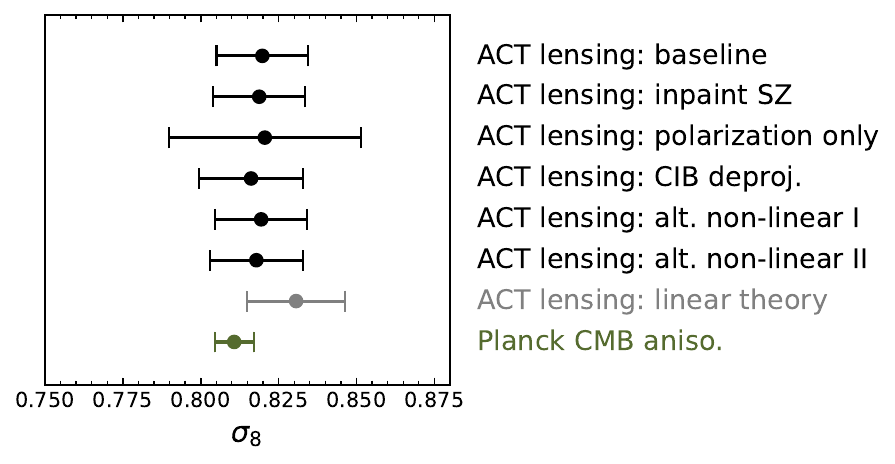}
    \caption{Marginalized posteriors for  $\sigma_8$ using variations of our ACT lensing analysis in combination with BAO data (black). The SZ inpainting method was our pre-unblinding result (see \citealt{Qu23}). We also show variations that use only polarization data and with an alternative CIB deprojection method for mitigating foregrounds. Constraints that use two alternative non-linear models from \cite{Mead2020} (with baryonic feedback) and from \cite{Casarini1,Casarini2} are also shown. The constraint that uses linear theory (gray) is not expected to agree perfectly, but the shift is small; together, these show that the details of the non-linear prescription do not matter significantly.}
    \label{fig:sigma8_data}
\end{figure}

\begin{table*}
\caption{Marginalized constraints on cosmological parameters in a consistent analysis of various weak lensing data-sets shown alongside CMB anisotropy (two-point) constraints. Throughout this work, we report the mean of the marginalized posterior and the 68\% confidence limit, unless otherwise mentioned.}
\label{tab:marg}
\begin{tabular*}{\textwidth}{ l  c  c  c  c}
\hline
\hline
Data                                                   & $\sigma_8$ & $S_8$ & $\Omega_{\rm m}$ & $H_0$ \\ 
                                                   &  &  &  & ($\text{km}\,\text{s}^{-1}\,\text{Mpc}^{-1}$) \\ \hline
Planck CMB aniso. (PR4 TT+TE+EE) + \texttt{SRoll2} low-$\ell$ EE & $0.811\pm0.006$ & $0.830\pm0.014$ & $0.314\pm0.007$ & $67.3\pm0.5$ \\ 
Planck CMB aniso. (+$A_{\rm lens}$ marg.) & $0.806\pm0.007$ & $0.817\pm0.016$ & $0.308\pm0.008$ & $67.8\pm0.6$ \\ 
ACT CMB Lensing + BAO & $0.820\pm0.015$ & $0.840\pm0.028$ & $0.315\pm0.016$ & $68.2\pm1.1$ \\ 
ACT+\Planck\ Lensing + BAO & $0.815\pm0.013$ & $0.830\pm0.023$ & $0.312\pm0.014$ & $68.1\pm1.0$ \\ 
ACT+\Planck\ Lensing (extended) + BAO & $0.820\pm0.013$ & $0.841\pm0.022$ & $0.316\pm0.013$ & $68.3\pm1.0$ \\ 
KiDS-1000 galaxy lensing + BAO & $0.732\pm0.049$ & $0.757\pm0.025$ & $0.323\pm0.034$ & $68.9\pm2.0$ \\ 
DES-Y3 galaxy lensing + BAO & $0.751\pm0.035$ & $0.773\pm0.025$ & $0.319\pm0.025$ & $68.7\pm1.5$ \\
HSC-Y3 galaxy lensing (Fourier) + BAO & $0.719\pm0.054$ & $0.766\pm0.029$ & $0.344\pm0.038$ & $70.2\pm2.3$ \\ 
HSC-Y3 galaxy lensing (Real) + BAO & $0.752\pm0.045$ & $0.760\pm0.030$ & $0.308\pm0.024$ & $68.0\pm1.5$ \\ 
\hline
\end{tabular*}
\end{table*}

Our companion papers, \cite{Qu23} and \cite{dr6-lensing-fgs}, provide detailed investigations of potential systematic effects in the lensing power spectrum measurement. In Figure \ref{fig:sigma8_data}, we perform inferences of $\sigma_8$ in combination with BAO for variations of the mass maps designed to test for our most significant systematic: astrophysical foregrounds. As explained in \cite{Qu23}, while our analysis was carefully blinded, a parallel investigation of the effect of masking and inpainting at the locations of SZ clusters led us to make a change in the pipeline post-unblinding; we find that this resulted in only a 0.03$\sigma$ shift in $\sigma_8$. Similarly, we find consistent results with an alternative foreground mitigation method (CIB deprojection; see \citealt{dr6-lensing-fgs} for details) and when using polarization data alone, where foreground contamination is expected to be significantly lower, although the uncertainties increase by a factor of two in the latter case. We also test the effect of using linear theory in the likelihood and find a 0.7$\sigma$ shift, which is expected but not so large as to raise concerns about our dependence on modeling non-linear scales. In addition, we replace our baseline non-linear modeling \citep{Mead2016} with alternative non-linear model I (\cite{Mead2020} with baryonic feedback) and non-linear model II \citep{Casarini1,Casarini2} and find negligible shifts. This robustness is expected from results from hydrodynamic simulations \citep{1910.09565,2011.06582}.

\subsection{Combination with \Planck\ lensing}

We compare and combine our lensing measurements with those made by the \Planck\ satellite experiment \citep{1807.06205}. We use the \NPIPE\ data release that re-processed \Planck\ time-ordered data with several improvements~\citep{2007.04997}. The \NPIPE\ lensing analysis~\citep{2206.07773} reconstructs lensing with  CMB angular scales from $100\leq\ell\leq2048$ using the quadratic estimator. Apart from incorporating around $8\%$ more data compared to the 2018 \Planck\ PR3 release, pipeline improvements were incorporated, including improved filtering of the reconstructed lensing field and of the input CMB fields (by taking into account the cross-correlation between temperature and $E$-polarization, as well as accounting for noise inhomogeneities; \citealp{2101.12193}). These raise the overall signal-to-noise ratio by around $20\%$ compared to  \Planck\ PR3~\citep{1807.06210}. Figure \ref{fig:Nell} shows a comparison of noise power between the \Planck\ PR3 lensing map and the \Planck\ \NPIPE\ lensing map.\footnote{The \NPIPE\ noise curve was provided by Julien Carron; private communication.}  The \NPIPE\ mass map covers 65\% of the total sky area in comparison to the ACT map which covers 23\%, but the ACT map described in Section \ref{sec:map} has a noise power that is at least two times lower, as seen in the same Figure.

Since the \NPIPE\ and ACT DR6 measurements only overlap over part of the sky, probe different angular scales, and have different noise and instrument-related systematics, they provide nearly independent lensing measurements. Thus, apart from comparing the two measurements, the consistency in terms of lensing amplitude and the $\scmbl \equiv \seightc$ lensing-only constraint as presented in \cite{Qu23} suggests that we may safely combine the two measurements at the likelihood level to provide tighter constraints. For the \NPIPE~ lensing measurements, we use the published \NPIPE\ lensing bandpowers, but use a modified covariance matrix to account for uncertainty in the normalization  in the same way as we do for ACT.\footnote{\url{https://github.com/carronj/planck_PR4_lensing}}
We compute the joint covariance between ACT and \NPIPE\ bandpowers using the same set of 480 full-sky FFP10 CMB simulations used by \NPIPE\ to obtain the \Planck\ part of the covariance matrix; see \cite{Qu23} for details. The resulting joint covariance indicates that the correlation coefficient between the amplitudes of the ACT and \Planck\ lensing measurements is approximately $18$\%. This is expected given the fact that although the ACT and \NPIPE\ data sets have substantially independent information, the sky overlap between both surveys means that there is still some degree of correlation between nearby lensing modes. 

The combination of ACT lensing, \Planck\ lensing, and BAO provides the following \apseightprecision\% marginalized constraint:
\begin{equation}
\sigma_8= \apseightmean \pm{\apseighterr} ,
\end{equation}
which is also consistent with the \Planck\ ~CMB anisotropy value $\sigma_8=0.811\pm0.006$ and the \WMAP\ $+$ ACT DR4 CMB anisotropy value $\sigma_8=0.819\pm0.011$.

\subsection{Comparison with galaxy surveys}

In order to place our constraints in the context of existing measurements, we use the most recently published galaxy weak lensing measurements from the Dark Energy Survey\footnote{\url{https://www.darkenergysurvey.org/}} (henceforth DES Y3), Kilo Degree Survey\footnote{\url{https://kids.strw.leidenuniv.nl/}} (henceforth KiDS-1000), and the Hyper Suprime-Cam Subaru Strategic Program\footnote{\url{https://hsc.mtk.nao.ac.jp/ssp/survey/}} (henceforth HSC-Y3). For each survey, we use the weak lensing shear two-point functions only; we do not include galaxy clustering or cross-correlations between galaxy overdensity and shear. While the three surveys provide similar statistical power, each has relative strengths and weaknesses: DES covers the greatest area  (approximately $5000\,\si{deg}^2$) with the lowest number density ($5.6$ galaxies per square arcminute),  while HSC-Y3 covers a relatively small area (approximately $416\,\si{deg}^2$) at much higher number density (15 galaxies per square arcminute). KiDS-1000 lies in the middle in both respects, and has the advantage of overlap with the VIKING survey \citep{edge13}, which provides imaging in five additional near infrared bands, enabling potential improvements in photometric redshift estimation. 

We use the published shear correlation function measurements and covariances from  DES Y3 and KiDS-1000, and Fourier-space and Real-space measurements from HSC-Y3. For our DES Y3 analysis we follow closely \citet{2105.13549,amon22,2105.13544}, using the same angular-scale ranges and modeling of intrinsic alignments, while for KiDS-1000 we follow closely \citet{2208.07179}, who reanalyzed galaxy weak lensing data sets, including KiDS-1000 after their initial cosmological analyses in \citet{2007.15633,2007.15632}. We follow the ``$\Delta\chi^2$ cut'' approach of \cite{2208.07179}, removing small-scale measurements to avoid marginalizing over theoretical uncertainty in the matter power spectrum due to baryonic feedback. For HSC-Y3, we show results from the HSC collaboration, who re-ran both their Fourier and Real-space analyses using the parameterization and priors shown in Table \ref{tab:params} in combination with galaxy BAO.  We provide further details of our analysis and comparison with published results in Appendix \ref{app:gal}.

Our results are shown in Figure \ref{fig:S8_bao} for two parameter combinations: (a) $S_8\equiv\seightg$ which is best constrained using galaxy weak lensing; and (b) the amplitude of matter fluctuations $\sigma_8$ alone. An interesting aspect of these results is that the $\sigma_8$ constraints from CMB lensing combined with BAO are significantly tighter than those from galaxy weak lensing shear combined with BAO. This difference arises from the different scale dependence of these two lensing observables, with galaxy lensing sensitive to much smaller scales than CMB lensing. We discuss this further in Appendix~\ref{app:tubes}. 

The CMB lensing measurements from ACT, \Planck, and SPTpol \citep{1910.07157}\footnote{The chains for this analysis were provided by the SPT collaboration; they have a slightly more conservative prior on $\ombh$ and do not include eBOSS DR16 LRG BAO, but this should not affect this comparison significantly. } are generally consistent with each other and with the \Planck\ CMB anisotropies. We find that for the $S_8$ parameter, the KiDS measurement, DES measurement, and HSC measurements (Fourier and Real-space) are lower than the \Planck\ CMB anisotropy constraint by roughly $2.6\sigma$, $2\sigma$, and 2 or $2.1\sigma$, respectively. With respect to the ACT+\Planck\ CMB lensing measurement, the KiDS, DES, HSC measurements are lower by $2.1\sigma$, $1.7\sigma$, $1.7$--$1.8\sigma$, respectively.

\begin{figure*}
    \centering
    \includegraphics[width=0.7\textwidth]{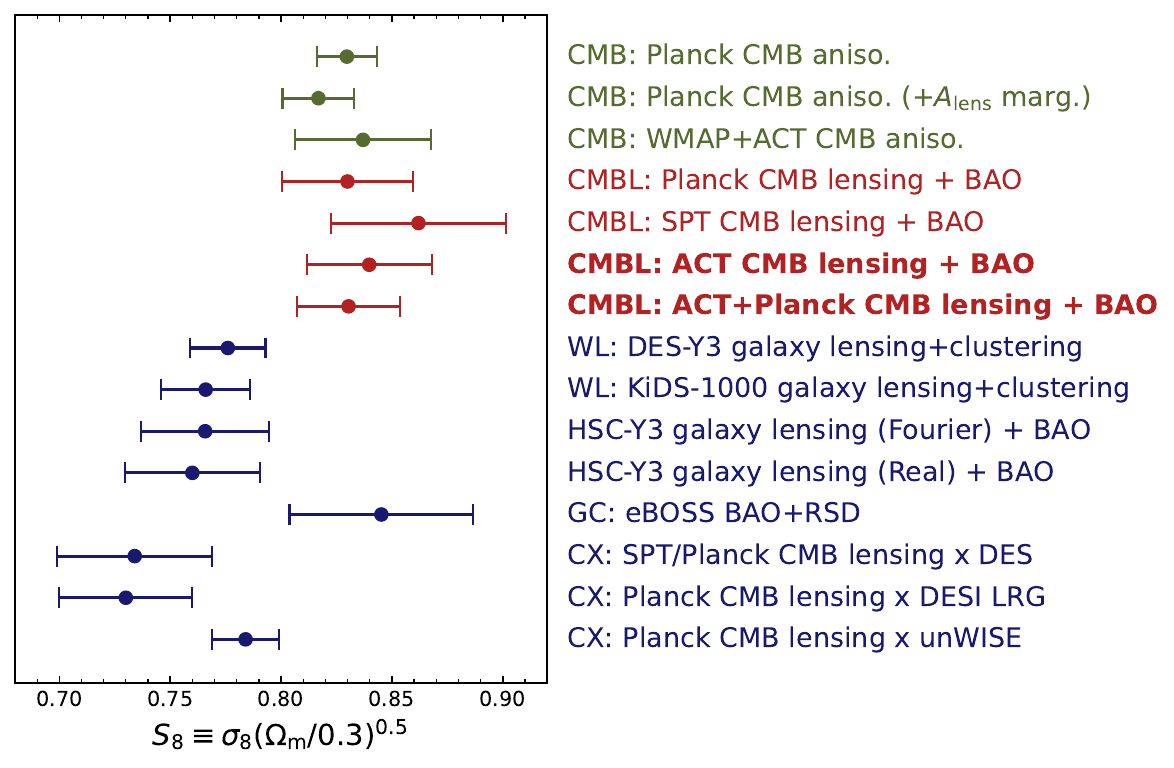}
    \caption{A comparison of $S_8=\seightg$ across multiple probes. We emphasize that the constraints in blue may not have been analyzed with consistent choices and priors but are values reported in the literature. Our CMB lensing measurements have relatively higher constraining power for $\scmbl=\seightc$ and, in combination with BAO, for $\sigma_8$; we refer the reader to Figure \ref{fig:S8_bao}. }
    \label{fig:s8_lss}
\end{figure*}

In Figure \ref{fig:s8_lss}, we show a more comprehensive comparison with a variety of large-scale structure probes. We caution that the probes shown in blue are not re-analyzed with consistent priors, but are drawn from the literature. We show constraints in the following categories:
\begin{enumerate}
    \item {\bf CMB:} These are CMB (two-point) anisotropy constraints, including our consistent reanalysis of \Planck\ PR4 CMB, with and without marginalization over $A_{\rm lens}$, and \WMAP+ACT DR4. This sets our expectation from the mainly primordial CMB view of the early universe.
    \item {\bf CMBL:} These are CMB lensing constraints with peak information from around $z=1$--$2$ from SPTpol \citep{1910.07157}, our re-analysis of \Planck\ \NPIPE\ \citep{2206.07773}, our baseline analysis of the new ACT DR6 CMB lensing mass map, and our combination of the latter with \Planck\ \NPIPE.
    \item {\bf WL: } These are large-scale structure measurements mainly driven by cosmic shear with optical weak lensing, but that may also include galaxy-galaxy lensing and galaxy clustering. We show constraints from the $\threecrosstwo$ DES-Y3 cosmology results \citep{2105.13549}, the KiDS-1000 $\threecrosstwo$ analysis \citep{2007.15632} and the HSC-Y3 galaxy lensing Fourier-space \citep{HSCY3Fourier} and Real-space analyses \citep{HSCY3Real}. 
    \item {\bf GC:} We show a constraint from galaxy clustering with the BOSS and eBOSS spectroscopic surveys, the final SDSS-IV cosmology analysis with BAO and RSD \citep{2007.08991} \footnote{This is obtained from the marginalized statistics of the chains \href{https://svn.sdss.org/public/data/eboss/DR16cosmo/tags/v1_0_1/mcmc/base/BAORSD_lenspriors/dist/base_BAORSD_lenspriors.margestats}{linked here}.}, which notably is consistent with CMB anisotropies. There have been several independent analyses of BOSS data using effective field theory (EFT) techniques. While some obtain consistent results \citep{2206.08327,2211.16794}, others such as \citep{2112.04515,2302.04414} obtain somewhat lower constraints on $S_8$ despite a large overlap in data.
    \item {\bf CX:} We show constraints derived from cross-correlations of CMB lensing from SPT and \Planck\ with various galaxy surveys. These include an SPT/\Planck\ CMB lensing cross-correlation with DES galaxies \citep{2203.12440}, a \Planck\ CMB lensing cross-correlation with DESI LRGs \citep{2111.09898}, and a \Planck\ CMB lensing cross-correlation with the unWISE galaxy sample \citep{2105.03421}. Interestingly, these constraints are lower than those from the \Planck\ CMB anisotropies and our CMB lensing measurement despite also involving CMB lensing mass maps.
\end{enumerate}

We find the general trend of CMB lensing measurements of large-scale structure (probing relatively higher redshifts and more linear scales) agreeing with the early-universe extrapolation from the CMB anisotropies. In contrast, there is a general trend of galaxy weak lensing probes finding lower inferences of structure growth.

\section{Cosmological constraints on expansion, reionization, and \LCDM~ extensions}

We now consider other parameters of interest both within \LCDM\, and in extended models.

\subsection{Hubble constant}

\begin{figure}
    \centering
    \includegraphics[width=\columnwidth]{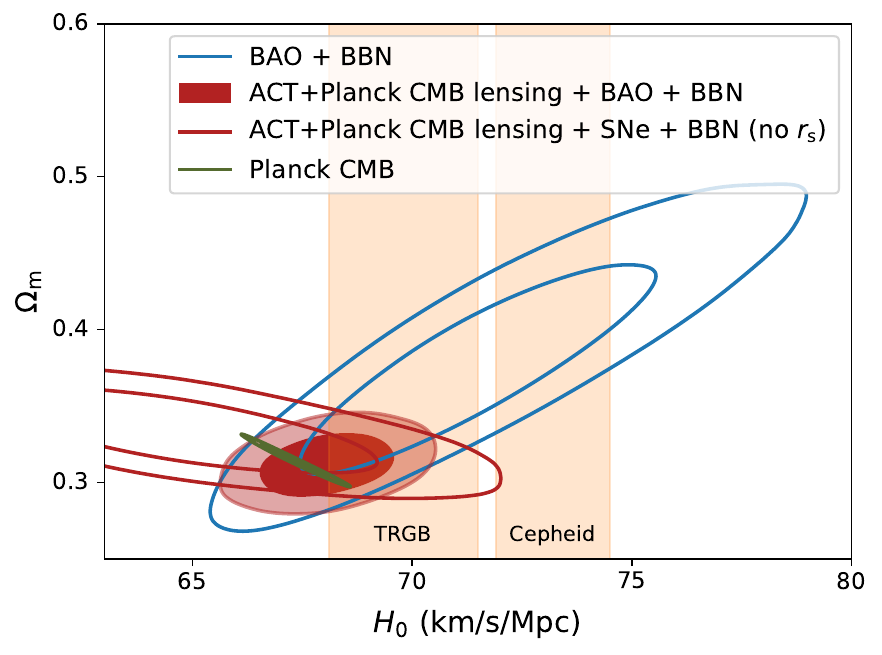}
    \caption{Hubble constant measurements with ACT CMB lensing. The red unfilled contours show a constraint that only utilizes $H_0$ information from the matter-radiation equality scale in contrast with indirect measurements that typically use the sound horizon scale. The addition of ACT lensing significantly improves the constraint from the combination of galaxy-only BAO and BBN (blue unfilled; using the BAO sample discussed in Section~\ref{sec:bao}), as can be seen in the red filled contours. The ACT lensing measurements are consistent with the low expansion rate inferred from \Planck~CMB anisotropies. They are in tension with the Cepheid-calibrated direct inference \citep{2112.04510} and are consistent with the TRGB-calibrated direct inference (\citealt{1907.05922}), whose 68\% c.l. bands are shown in orange.}  
    \label{fig:H0_bao}
\end{figure}

\begin{figure}[t]
    \centering   \includegraphics[width=\columnwidth]{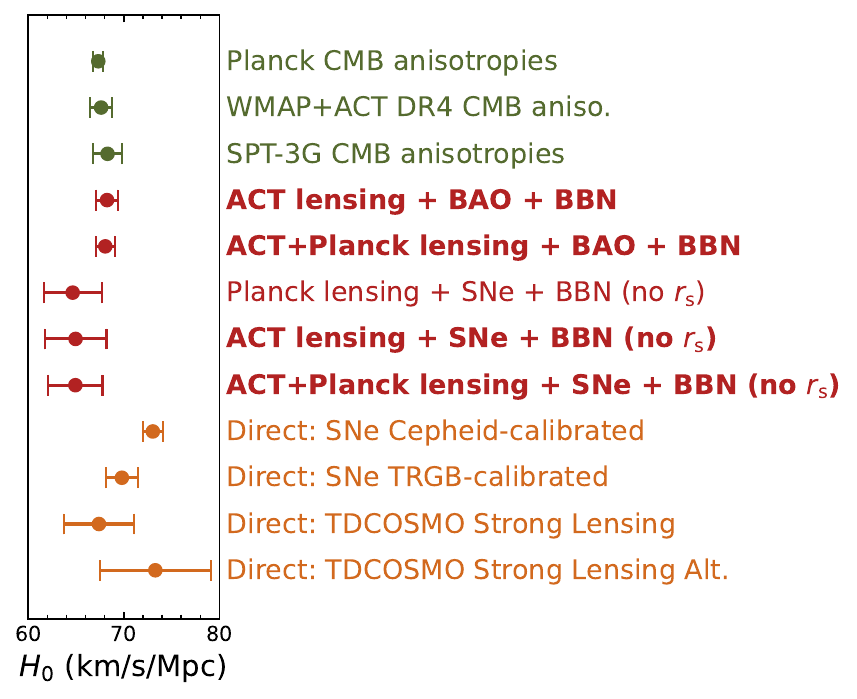}
    \caption{Marginalized posteriors for the Hubble constant from ACT lensing (red). We show constraints both from the combination with BBN and BAO (which depends on the sound horizon $r_{\rm s}$) and on a combination with BBN \texttt{Pantheon+} supernovae (no $r_{\rm s}$ dependence). We also show various CMB anisotropy measurements that are primarily an early-universe extrapolation (green), and direct inferences of the Hubble constant (orange) from the local universe.}
    \label{fig:H0}
\end{figure}

Our DR6 CMB lensing measurements also provide independent constraints on the Hubble constant. The first method by which our lensing results can contribute to expansion-rate measurements is via the combination with galaxy BAO data. As seen in Figure \ref{fig:H0_bao}, if we consider galaxy BAO observations without CMB lensing (but with a BBN prior on the baryon density, which contributes to calibrating the BAO scale via the sound horizon $r_{\rm d}$), the constraints on $H_0$ are still quite weak (empty blue contours); this is due to an extended degeneracy direction between $H_0$ and $\Omega_{\text{m}}$. However, the CMB lensing power spectrum constraints exhibit a degeneracy direction between $H_0$ and $\Omega_{\text{m}}$ that is nearly orthogonal to the BAO constraints. Therefore, the combination of $r_d$-calibrated galaxy BAO and CMB lensing allows degeneracies to be broken, and tight constraints to be placed on the Hubble constant, as shown in Figures \ref{fig:H0_bao} and \ref{fig:H0}. In particular, from the combination of ACT CMB lensing, galaxy BAO, and a BBN prior, we obtain the constraint:
\begin{equation}
H_0= \ahmean \pm{\aherr} \, {\rm km}\,{\rm s}^{-1}\,{\rm Mpc}^{-1} .
\end{equation}
Similarly, using the combination of ACT and \Planck\ CMB lensing together with BAO and a BBN prior, we obtain
\begin{equation}
H_0= \aphmean \pm{\apherr} \, {\rm km}\,{\rm s}^{-1}\,{\rm Mpc}^{-1} .
\end{equation}
Both constraints are consistent with $\Lambda$CDM-based Hubble constant inferences from the CMB and large-scale structure, and with the tip of the red-giant branch (TRGB)-calibrated local distance ladder measurements from \cite{1907.05922}, but are in approximately $3.4\sigma$ tension with the local distance-ladder measurements from \texttt{SH0ES} of $H_0= 73.04 \pm{1.04} \, {\rm km}\,{\rm s}^{-1}\,{\rm Mpc}^{-1}$ \citep{2112.04510}. 

We expect the above constraints to be primarily derived from the angular and redshift separation subtended by the BAO scale\footnote{While we have not proven this, it has been shown that if data sets that calibrate the BAO scale (such as BBN) are included, the BAO feature has most constraining power and dominates the large-scale structure (LSS) inference of the Hubble constant \citep{2204.02984}.}, which is set by the comoving sound horizon at the baryon drag epoch, $r_{\rm d}$ \citep{EisensteinHu1998}. The majority of current CMB and LSS constraints that are in tension with local measurements from \texttt{SH0ES} derive from this sound horizon scale\footnote{Here, we do not make a careful distinction between the sound horizon scale relevant for LSS ($r_{\rm d}$) and CMB ($r_{\rm s}$) observations, although to be precise these are defined at the baryon drag epoch and at photon decoupling, respectively.}. This fact has motivated theoretical work to explain the tension by invoking new physics that decreases the physical size of the sound horizon at recombination by approximately $10\%$ (e.g.,~\citealp{1811.00537,1908.03663}). This situation motivates new measurements of the Hubble constant that are derived from a different physical scale present in the large-scale structure, namely, the matter-radiation equality redshift and scale (with comoving wave-number $k_{\rm eq}$) which sets the turn-over in the matter power spectrum.

Over the past two years, several measurements of the Hubble constant that rely on the matter-radiation equality information and are independent of the sound horizon scale have been performed, giving results that are consistent with values of $H_0$ derived from the sound horizon scale (e.g., \citealp{2007.04007,2204.02984}). Here, we repeat the analysis method used in \cite{2007.04007} and applied to \Planck\ data to obtain sound-horizon-independent $H_0$ measurements from both ACT and \Planck\ CMB lensing data and their combination. In particular, we combine CMB lensing power spectra -- which are sensitive to the matter-radiation equality scale and hence, in angular projection, $\Omega_{\text{m}}^{0.6}h$  -- with uncalibrated supernovae from \texttt{Pantheon+} \citep{2202.04077}, which independently constrain $\Omega_{\text{m}}$ through the shape of the redshift-apparent brightness relation. Here, `uncalibrated' refers to the fact that the absolute magnitudes of the supernovae have not been calibrated, e.g. with Cepheid variables or the TRGB technique, such that only information from the relative (not absolute) distance-redshift relation is included. This combination, along with suitable prior choices as in \cite{2007.04007}, allows us to constrain $H_0$. For the following $r_{\rm s}$-independent constraints that exclude BAO, we sample in $H_0$ instead of $\thetamc$ and impose a prior of $\omegam=0.334 \pm 0.018$ corresponding to the \texttt{Pantheon+} \citep{2202.04077} measurement. With this approach, we obtain from ACT lensing\footnote{{The reader may wonder about the difference -- $10\,\text{km}\,\text{s}^{-1}\,\text{Mpc}^{-1}$ lower here -- with the value determined by~\citet{2007.04007}, which used \Planck \ + Pantheon. We believe that the change from Pantheon to Pantheon+ is, to a significant extent, responsible for this difference -- the Pantheon+ $\Omega_{\rm m}$ is 13\% higher than Pantheon, which lowers $h$ in this analysis; this also matches what was found in \cite{2204.02984} using BOSS, \emph{Planck} lensing, and Pantheon+.}} 
\begin{equation}
H_0= \ahmeanalt \pm{\aherralt} \, {\rm km}\,{\rm s}^{-1}\,{\rm Mpc}^{-1} .
\end{equation}
With the combination of both \Planck\ and ACT lensing, we have
\begin{equation}
H_0= \aphmeanalt \pm{\apherralt} \, {\rm km}\,{\rm s}^{-1}\,{\rm Mpc}^{-1} .
\end{equation}
As seen in Figure \ref{fig:H0_bao}, this constraint is also low (at 2.7$\sigma$ significance) compared to the \texttt{SH0ES} result, although it derives from different early-universe physics than the standard BAO or CMB Hubble constant measurements.

In Figure \ref{fig:H0}, we show both our marginalized $r_{\rm s}$-independent Hubble constant constraints and those from combination with BAO against a compilation of various other indirect and direct constraints. We show in green measurements from the power spectra of the CMB anisotropies including those described in Appendix~\ref{app:cmb}: i.e., from \Planck\ (the combination including \NPIPE), from ACT DR4 (the combination with {\it WMAP}), as well as the SPT-3G CMB measurement \citep{2101.01684}. Among direct measurements, we show the TDCOSMO strong-lensing time-delay measurement with marginalization over lens profiles \citep{2007.02941}, an alternative TDCOSMO measurement with different lens-mass assumptions \citep{2007.02941}, the TRGB-calibrated supernovae measurement \citep{1907.05922}, and the Cepheid-calibrated \texttt{SH0ES} supernovae measurement \citep{2112.04510}.

The consistency of our $r_{\rm s}$-independent and $r_{\rm s}$-based inferences of $H_0$ provides significant support to the idea that the standard $\Lambda$CDM model accurately describes the pre-recombination universe.  Although $r_{\rm s}$-independent $H_0$ inferences become less constraining in many extended models~\citep{2208.12992}, the comparison of $r_{\rm s}$-based and $r_{\rm s}$-independent constraints is nevertheless a non-trivial null test for $\Lambda$CDM (e.g.,~\citealp{2112.10749,2204.02984,2212.04522}), which the model currently passes.  The consistency observed here does not provide support to models that attempt to increase the inferred value of $H_0$ via changes to sound horizon physics.

\subsection{Neutrino mass}\label{sec:mnu}

Observations showing neutrinos oscillate from one flavor to another require these particles to have mass. This is of considerable consequence for particle physics since plausible mechanisms for generating neutrino masses require physics beyond the Standard Model (BSM).\footnote{In some scenarios, measured neutrino masses can map directly on to parameters of BSM Lagrangians like the Majorana phases. See \cite{PhysRevLett.130.051801} for recent Majorana neutrino search results from KamLAND-Zen.} Cosmological surveys will provide important constraints in this sector \citep{1509.07471,1610.02743,1611.00036,1808.07445}. While neutrino oscillation experiments measure the differences of squared mass $\Delta m^2_{1,2}$ and $|\Delta m^2_{3,2}|$ between pairs of the three mass eigenstates, they do not tell us the absolute scale or sum of the masses. However, given the measured mass-squared differences, we know that the sum of neutrino masses $\mnu$ must be {\emph{at least}} $58\,\si{meV}$ for a normal hierarchy (two masses significantly smaller than the third) and $100\,\si{meV}$ for an inverted hierarchy (two masses significantly higher than the third). This sets clear targets for experiments that aim to measure the overall mass scale.

Direct experiments like KATRIN \citep[see recent results in][]{2105.08533} that make observations of tritium beta decay will constrain $\mnu$ to below $200\,\si{meV}$ (90\% c.l.) over the next decade.\footnote{The proposed Project 8 could reach a constraint of 40 meV (90\% c.l.) \citep{10.1103/PhysRevD.80.051301,1703.02037,2012.14341}, which would allow for a valuable comparison of a direct measurement with a cosmological measurement even for relatively low mass scales.} Cosmological observations sensitive to the total matter power spectrum on the other hand have already provided stronger constraints (e.g., \citealp{1807.06209}), albeit contingent on assumptions in the \LCDM\ standard model of cosmology. As the universe expands, neutrinos cool and become non-relativistic at redshifts $z\simeq 200 \left(\sum m_{\nu} / 100\,\si{meV}\right)$. On scales larger than the neutrino free-streaming length, neutrinos cluster and behave like CDM. On smaller scales, their large thermal dispersion suppresses their clustering while their energy density contributes to the expansion rate, also causing the growth of CDM and baryon perturbations to be suppressed. Thus the net effect is a suppression of the overall (dark-matter dominated) matter power spectrum on scales smaller than the neutrino free-streaming length. 

Cosmological observations do not resolve the scale-dependence very well currently, so the dominant signal we look for is an overall suppression of the matter power spectrum relative to that extrapolated from the early-time cosmology measured from the primary CMB anisotropies.  Since massive neutrinos suppress the matter power spectrum, and since the CMB lensing power spectrum is a line-of-sight projected integral over this power spectrum, CMB lensing is an excellent probe of massive neutrinos.\footnote{This suppression is however degenerate with the physical matter density $\omegam h^2$ and hence it is crucial to incorporate BAO data that helps break this degeneracy \citep{1506.07493}.}\\

\begin{figure}[t]
    \centering
    \includegraphics[width=\columnwidth]{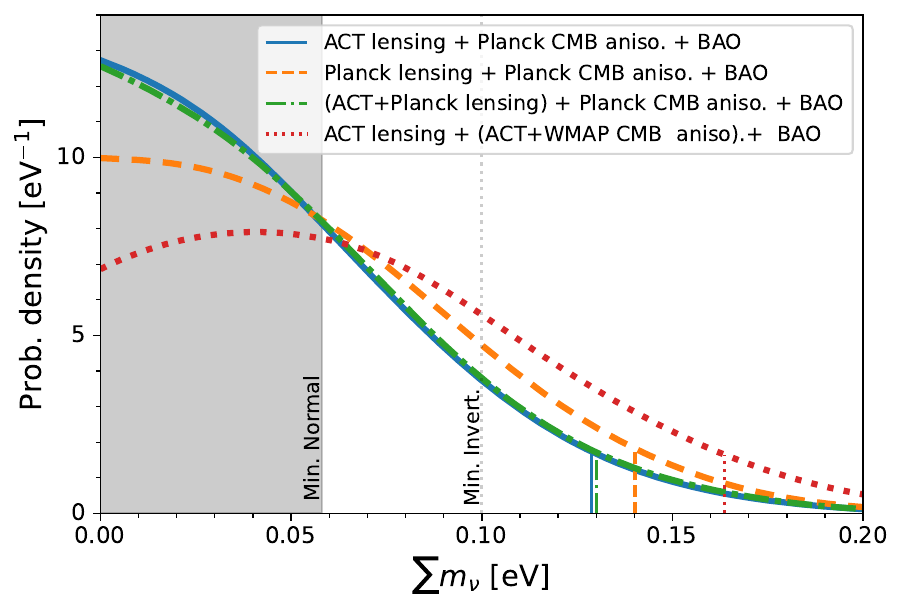}
    \caption{Marginalized posterior probability densities for the sum of neutrino masses from ACT CMB lensing. The vertical lines show the corresponding 95\% confidence limits. All constraints here include BAO data, primary CMB anisotropy data, and optical depth information from \Planck\ polarization in addition to CMB lensing. For our baseline constraints, we use CMB anisotropy data from \Planck, but we also show in the red dotted curve the constraint obtained when using ACT DR4+\WMAP\ for the CMB anisotropies. With ACT, the posterior is peaked at higher neutrino masses. The minimal sum of masses expected from oscillation experiments in a normal hierarchy and inverted hierarchy are shown as solid gray and dotted gray vertical lines, respectively. }
    \label{fig:mnu}
\end{figure}

We combine our ACT lensing measurement with BAO and CMB anisotropies to obtain constraints on $\mnu$ in a seven-parameter model (see Table \ref{tab:params})\footnote{Following the arguments in \cite{10.1016/j.physrep.2006.04.001} and \cite{1612.00021}, we consider a degenerate combination of three equally massive neutrinos when varying $\mnu$.}. The lensing measurement together with BAO provides a handle on the amplitude of matter fluctuations at late times and the CMB anisotropies provide an anchor in the early universe that measures primordial fluctuations. The sum of neutrino masses can then be inferred through relative suppression in the matter power at late times; we show our results in Figure \ref{fig:mnu}. Our baseline constraint uses ACT lensing with \Planck\ CMB anisotropies (as well as galaxy BAO and optical depth information from the \texttt{SRoll2} re-analysis of the \Planck\ data; see~\citealt{pagano/etal:2020} and Appendix~\ref{app:cmb}):
\begin{equation}
\mnu <  \amnu\, \text{eV;~ 95\%~ c.l.}
\end{equation}
This can be compared to the constraint we obtain when replacing the ACT lensing information with \Planck\ NPIPE lensing of $\mnu <  0.14\,\text{eV;~ 95\%~ c.l.}$. Combining the ACT and \Planck\ lensing measurements, we have
\begin{equation}
\mnu <  \apmnu\, \text{eV;~ 95\%~ c.l.} 
\end{equation}
The combination of ACT and \Planck\ lensing gives a similar bound to ACT alone despite improving the Fisher information; this is likely due to the lower value of $\sigma_8$ preferred by the combination.  We also note that analyses that use \Planck\ PR3 CMB anisotropy data, including \Planck\ PR3 lensing \citep{1807.06210,1807.06209} and eBOSS galaxy clustering \citep{2007.08991}, obtain a similar constraint of $\mnu <  0.12\, \text{eV;~ 95\%~ c.l.}$. At face value this suggests that adding ACT lensing does not bring new information. However, we note that variations in the \Planck\ CMB anisotropy data have an impact on this upper limit. In particular, \Planck\ PR3 CMB power spectra prefer a high fluctuation in the lensing peak smearing, which tends to lead to a preference for lower neutrino masses and a tighter bound that does not need to be commensurate with the Fisher information in the data set\footnote{ Indeed, the constraint we obtain using \Planck\ PR3 CMB
anisotropies is tighter; for the ACT+\Planck\ lensing combination with the extended multipole range, the constraint tightens from $\mnu <  0.13$ to $\mnu <  0.12\, \text{eV;~ 95\%~ c.l.}$ when switching from PR4 to PR3.}. This effect is reduced with the \Planck\ PR4 anisotropies (\Planck\ PR4 CMB + BAO alone yields $\mnu <  0.16\, \text{eV;~ 95\%~ c.l.}$) used here and as a net result, even though we use more data, we recover a similar bound. We also obtain an alternative constraint that swaps the \Planck~ CMB anisotropies with measurements from {\it WMAP} and ACT DR4. In this case, the posterior peak shifts to higher values and the bound weakens to

\begin{equation}
\mnu <  0.16 ~{\rm eV;~ 95\%~ c.l.}
\end{equation}

The constraint on the optical depth to reionization is an important input in these inferences since the suppression of matter power is obtained relative to the measured early-universe fluctuations which are screened (and suppressed) by the reionization epoch \citep{10.1103/PhysRevD.55.1822}. As noted above, our baseline constraints use an updated analysis of low-$\ell$ \Planck\ polarization data from \texttt{SRoll2}, but we also obtain a constraint on $\mnu$ using a much more conservative Gaussian prior on the optical depth of $\tau=0.06\pm 0.01$:
\begin{equation}
\mnu <  0.15\,\text{eV;~ 95\%~ c.l.}
\end{equation}

\subsection{Curvature and dark-energy density}

Spatial flatness of the universe is a key prediction of the inflationary paradigm underpinning the standard model of cosmology. There has been a suggestion that the \Planck\ CMB anisotropies prefer a closed universe (with curvature parameter $\omegak<0$, where $\omegak = 1 - \omegam - \omegal$), driven entirely by the moderately high lensing-like peak smearing in \Planck\ measurements of the CMB anisotropies  \citep{1911.02087}. It should be noted that this preference for negative curvature density weakens in the recent \Planck\ \NPIPE\ re-analysis of CMB anisotropies \citep{rosenberg/etal:2022}. An independent measurement from ACT DR4+\WMAP\ \citep{Aiola,Choi} is consistent with zero spatial curvature. The combination of BAO and primary CMB data also strongly favors a flat universe.

Nevertheless, we revisit these constraints using CMB data alone. The primary CMB anisotropies alone do not constrain curvature due to a ``geometric degeneracy'' \citep{10.1086/185100,9807103} that is broken with the addition of lensing information~\citep{9805294}. Since the ACT and \Planck\ lensing measurements are consistent with the flat \LCDM\ prediction, we expect a zero curvature preference to return when including the full lensing information in the mass map, as also seen with \Planck\ data in \cite{1807.06210}.  We therefore perform inference runs in a \LCDM+$\omegak$ extension. 

\begin{figure}[t]
    \centering \includegraphics[width=\columnwidth]{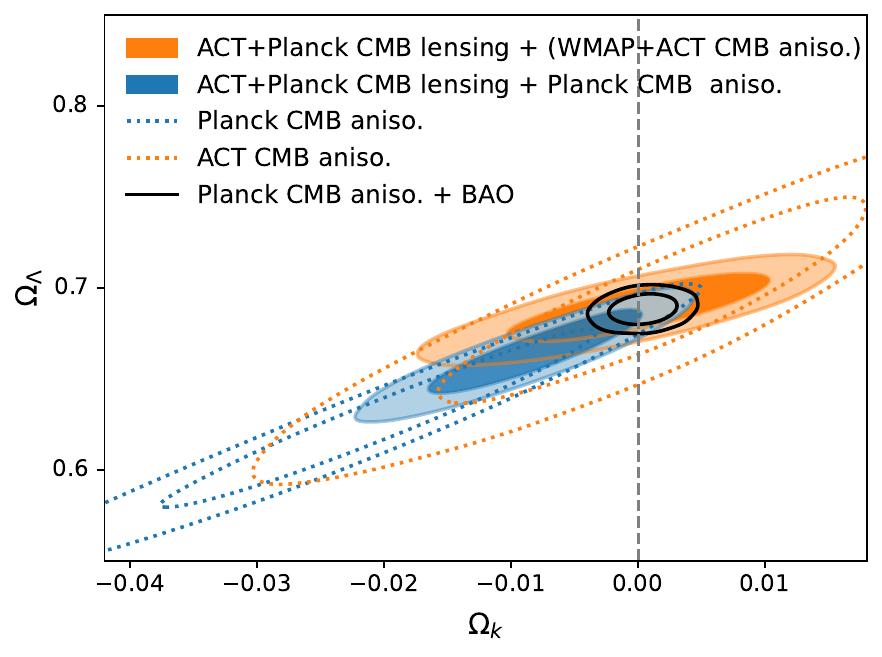}
    \caption{Constraints on spatial curvature and the dark-energy density from ACT lensing in a \LCDM+$\Omega_{\rm k}$ model. The dotted contours show constraints in this plane from CMB anisotropies from ACT DR4 or \Planck; these suffer from a geometric degeneracy that is weakly broken with the lensing information in the smearing of the CMB acoustic peaks. Including the full lensing information from our ACT lensing power spectrum significantly reduces this degeneracy and provides: (a) consistency with zero spatial curvature; and (b) a high-significance detection of a dark-energy component from the CMB alone. Addition of BAO data significantly tightens the constraint around zero spatial curvature.}
    \label{fig:omk}
\end{figure}

We show our results in Figure \ref{fig:omk} in the $\omegal$--$\omegak$ plane. The addition of ACT+\Planck\ lensing data to \Planck\ CMB anisotropies gives
\begin{equation}
-0.019 < \omegak <  0.002 ~{\rm ~ 95\%~ c.l.},
\end{equation}
and replacing the CMB anisotropies with those from {\it WMAP}+ACT DR4 gives
\begin{equation}
-0.013 < \omegak <  0.013 ~{\rm ~ 95\%~ c.l.}
\end{equation}
Both are consistent with spatial flatness. We note that the above constraints only use CMB data and can be equivalently seen as constraining the energy density due to a cosmological constant. For example, as done in \cite{1105.0419}, we have from CMB data alone, and limiting to ACT lensing alone with {\it WMAP} + ACT DR4 CMB anisotropies
\begin{equation}
\omegal = 0.68\pm 0.01 .
\end{equation}
with the accompanying curvature constraint of $-0.016 < \omegak <  0.012 ~({\rm 95\%~ c.l.})$. While the combination of CMB lensing and CMB anisotropies provides constraints consistent with spatial flatness, we note that combining BAO and CMB anisotropies provides a much tighter constraint. For example, with galaxy BAO and \Planck\ CMB anisotropies, the curvature density is constrained to $-0.003 < \omegak <  0.004 ~{\rm ~ 95\%~ c.l.}$ (see Figure \ref{fig:omk}). This is not improved significantly with further combination with CMB lensing, but the consistency with flatness from the combination of CMB anisotropy and CMB lensing provides an important cross-check.

\subsection{Reionization}

In the above analyses, we have used low-$\ell$ \Planck\ polarization data to break a degeneracy of the late-time matter fluctuation amplitude with the optical depth to reionization $\tau$. This degeneracy arises from the fact that in order to probe effects that change the late-time matter fluctuation amplitude, one must measure and extrapolate from the primordial fluctuations (with amplitude $A_s$) encoded in the CMB anisotropies. These anisotropies are, however, screened and suppressed during the reionization epoch; the power spectra scale as $A_{\text{s}}^2e^{-2\tau}$ on intermediate and small angular scales. The low-$\ell$ CMB polarization `reionization bump' provides the required independent information on the optical depth $\tau$ to break this degeneracy. 

Measuring polarization at low-$\ell$ (on large angular scales) is, however, challenging due to a variety of instrumental and astrophysical systematic effects. It is therefore interesting to turn the question around and ask whether we can infer the optical depth independently from low-$\ell$ polarization by comparing the CMB lensing-inferred late-time matter fluctuation amplitude with the primordial fluctuations in the CMB anisotropies~\citep{1502.01591,1502.01589}. This requires choosing and fixing a model to perform the extrapolation from the CMB anisotropies to the late-time lensing observations; we choose our baseline \LCDM~ model with $\mnu=0.06\,{\rm eV}$ and the six cosmological parameters varied (see Table \ref{tab:params}). Using ACT+\Planck\ lensing, BAO, and \Planck\ CMB anisotropies (excluding low-$\ell$ polarization), we obtain within this model
\begin{equation}
\tau = 0.074 \pm 0.014,
\end{equation}
and using {\it WMAP}+ACT DR4 CMB anisotropies instead of \Planck
\begin{equation}
\tau = 0.058\pm 0.015.
\end{equation}
These constraints on the optical depth to reionization independent of low-$\ell$ CMB polarization data are consistent with the value $\tau=0.059\pm0.006$ obtained from the \texttt{SRoll2} low-$\ell$ polarization analysis \citep{pagano/etal:2020}.

\section{Data products}

This article is accompanied by a release of the likelihood software required to reproduce the ACT cosmological constraints. The CMB lensing mass map will also be made publicly available. In this section, we provide details of these data products.

\subsection{Using the mass map}

The mass map is provided as a \fits~ file containing the spherical harmonic modes $\kappa_{LM}$ of the map in a format suitable to be loaded by software like \texttt{healpy}. These can be projected onto desired pixelization schemes, e.g., the \texttt{HEALPix} equal-area pixel scheme, but we note that the map is in an Equatorial coordinate system as opposed to the Galactic coordinate system of \Planck\ maps. The map has been top-hat filtered to remove unreliable scales outside multipoles $40 < L < 3000$; this filter must be forward-modeled in any real-space or stacking analysis. This baseline map is a minimum-variance combination of CMB temperature and polarization information with foreground mitigation through profile hardening, but we also provide variants as described in Section \ref{sec:clusters}.

We provide the analysis mask that was used when preparing the input CMB maps. When using the mass map for cross-correlations, it is often necessary to deconvolve the mask, e.g., using the MASTER algorithm \citep{Hivon}. We caution that this procedure is not exact in the case of CMB lensing mass maps, since they are quadratic in the input CMB maps. An approximate way to account for this is to use the square of the analysis mask in software packages like \texttt{NaMaster} \citep{Namaster} that implement the MASTER algorithm.

Regardless of the approach used, we strongly encourage users of the mass map to use the provided simulations to test their pipeline for (and estimate) a possible multiplicative transfer function, especially in situations where the area involved in the cross-correlation is significantly smaller than the ACT mass map. We provide both simulated reconstruction maps as well as the input lensing convergence maps for this purpose.

\subsection{Cluster locations, astrophysical foregrounds, and null maps}\label{sec:clusters}

The standard quadratic estimator we have used \citep{HuOk} suffers from a known issue at the location of massive clusters; the reconstruction becomes biased low in these regions due to higher-order effects \citep{HDV}. For this reason, we provide a mask of SZ clusters to avoid when stacking. Cross-correlations with most galaxy samples should not be affected.

We also provide lensing reconstructions run on simulations that contain the Websky implementation of extragalactic foregrounds~\citep{2001.08787}. We encourage users of the mass maps to implement a halo-occupation-distribution (HOD) for their galaxy sample of interest into the Websky halo catalog so as to test with these simulations for any possible residual foreground bias. These simulations can also be used to test for possible effects due to correlations between the mask and large-scale structure (see, e.g., \citealt{2302.05436}). For similar purposes, we provide a suite of null maps (e.g., lensing reconstruction performed on the difference of 90\,GHz and 150\,GHz maps) that can be cross-correlated with large-scale structure maps of interest. We additionally provide the following variants of the lensing map that can be used to assess foreground biases: (a) one that utilizes only CMB polarization information (b) one that utilizes only CMB temperature information and (c) one that uses an alternative foreground mitigation procedure involving spectral deprojection of the CIB.

\subsection{Likelihood package and chains}

We provide the bandpowers of the lensing power spectrum measurement, a covariance matrix, and a binning matrix that can be applied to a theory prediction. We also provide a Python package that contains a generic likelihood function as well as an implementation for the \texttt{Cobaya} Bayesian inference framework. We provide variants corresponding both to the pre-unblinding `baseline' multipole range of $40 < L < 763$ and the `extended' multipole range of $40 < L < 1300$, set after unblinding.

\section{Conclusion and Discussion}

We have used ACT CMB data from 2017 to 2021 to provide a new view of large-scale structure through gravitational lensing of the CMB, providing a high-fidelity wide-area mass map covering $9400\,\si{deg}^2$ to the community for further cross-correlation science. Through a study of the power spectrum of this mass map, measured in \cite{Qu23}, in combination with BAO data, we find that the amplitude of matter fluctuations $\sigma_8$ is  consistent (at \aseightprecision\% precision) with the expectation from the \LCDM\ model fit to measurements of the CMB anisotropies from \Planck\ that probe mainly the early universe. We find that a consistent re-analysis of galaxy weak lensing (cosmic shear) data with identical prior choices shows all three of DES, HSC, and KiDS to be lower than \Planck\ anisotropies at varying levels ranging from $2$--$2.6\sigma$ and lower than our ACT+\Planck\ lensing measurement at varying levels ranging from $1.7$--$2.1\sigma$. We find a CMB lensing-inferred value of the Hubble constant $H_0$ consistent with \Planck\ $\Lambda$CDM and inconsistent with Cepheid-calibrated supernovae; this persists even when analyzing a variant of our measurement that does not derive information from the sound horizon. Our joint ACT+\Planck\ lensing constraint on the sum of neutrino masses $\mnu <  \apmnu\,\text{eV (95\% c.l.)}$ and $\mnu <  \apmnuh\,\text{eV (99\% c.l.)}$ provides a robust measurement that relies on mostly linear scales. With CMB data alone, informed by ACT lensing, we find that the universe is consistent with spatial flatness and requires a dark energy component.

We have only considered a subset of interesting model extensions here. Our publicly released likelihoods encapsulate linear scales of the total matter density field primarily over the redshift range $z=0.5$--$5$. A variety of follow-up investigations will be of interest, including those that combine with galaxy lensing and clustering covering a range of redshifts and scales, possibly fitting these measurements jointly with models that look for non-standard dark matter physics and modifications of general relativity. An exciting near-term prospect is an exclusion of the inverted hierarchy of massive neutrinos; for example, improved BAO data from the ongoing Dark Energy Spectroscopic Instrument (DESI; \citealp{1611.00036}) will significantly reduce the degeneracy of our $\mnu$ measurement with the matter density $\omegam$ \citep{1509.07471}. 

The publicly released mass maps can be used for a variety of cross-correlations; those with galaxy surveys, for example, can produce improved constraints on local primordial non-Gaussianity $f_{\rm NL}$ \citep{1710.09465,2210.01049}, as well as constraints on the amplitude of structure as a function of redshift $\sigma_8(z)$ \citep[e.g.,][]{2111.09898}. The mass maps can be combined with measurements of the thermal and kinetic Sunyaev--Zeldovich effects along with X-ray measurements to study the thermodynamics of galaxy formation and evolution by supplementing electron pressure, density, and temperature measurements with gravitational mass on arcminute scales \citep{1705.05881,2023JCAP...03..039B}. They can also be used to study the non-linear universe, providing an unbiased view of the distribution of voids and filaments (e.g., \citealt{1709.02543,1911.08475}).

ACT completed observations in 2022, but several possibilities lie ahead for significantly improved mass maps and cosmological constraints.  In particular, we will explore the fidelity of roughly 50\% of ACT data collected (mostly during the day-time) that was not used in this analysis. Data at lower frequencies and at 220\,GHz can be used to enhance the foreground cleaning, which in combination with hybrid mitigation strategies \citep{2111.00462} may allow us to use higher multipoles in the CMB lensing reconstruction. Other areas of exploration include: (a) optimal filtering of ACT maps that accounts for noise non-idealities \citep{1909.02653}; (b) CMB-map-level combination with \Planck~ data; (c) improved accuracy and precision of the lensing signal at the location of galaxy clusters \citep{HDV}; and (d) improved compact-object treatment allowing for less aggressive masking of the Galaxy, thus enabling larger sky coverage of the mass map.

Looking further ahead, the Simons Observatory~\citep{1808.07445}, under construction at the same site as ACT, will significantly improve the sensitivity of CMB maps. This will enable sub-percent constraints on the amplitude of matter fluctuations and a wide variety of cosmological and astrophysical science goals.

\section*{Acknowledgments}

We are grateful to Marika Asgari, Federico Bianchini, Julien Carron, Chihway Chang, Antony Lewis, Emily Longley, Hironao Miyatake, Jessie Muir, Luca Pagano, Kimmy Wu and Joe Zuntz for help with various aspects of the external codes and data-sets used here. We are especially grateful to Xiangchong Li and the HSC team for making a consistent analysis of their Y3 results available to us.  Some of the results in this paper have been derived using the \texttt{healpy}~\citep{Healpix2} and \texttt{HEALPix}~\citep{Healpix1} packages. This research made use of \texttt{Astropy},\footnote{http://www.astropy.org} a community-developed core Python package for Astronomy \citep{astropy:2013, astropy:2018}. We also acknowledge use of the \texttt{matplotlib}~\citep{Hunter:2007} package and the Python Image Library for producing plots in this paper, use of the Boltzmann code \texttt{CAMB}~\citep{CAMB} for calculating theory spectra, and use of the \texttt{GetDist} \citep{1910.13970}, \texttt{Cobaya} \citep{2005.05290} and \texttt{CosmoSIS} \citep{1409.3409} software for likelihood analysis and sampling.  We acknowledge work done by the Simons Observatory Pipeline and Analysis Working Groups in developing open-source software used in this paper.

Support for ACT was through the U.S.~National Science Foundation through awards AST-0408698, AST-0965625, and AST-1440226 for the ACT project, as well as awards PHY-0355328, PHY-0855887 and PHY-1214379. Funding was also provided by Princeton University, the University of Pennsylvania, and a Canada Foundation for Innovation (CFI) award to UBC. ACT operated in the Parque Astron\'omico Atacama in northern Chile under the auspices of the Agencia Nacional de Investigaci\'on y Desarrollo (ANID). The development of multichroic detectors and lenses was supported by NASA grants NNX13AE56G and NNX14AB58G. Detector research at NIST was supported by the NIST Innovations in Measurement Science program. 

Computing was performed using the Princeton Research Computing resources at Princeton University, the Niagara supercomputer at the SciNet HPC Consortium and the Symmetry cluster at the Perimeter Institute. SciNet is funded by the CFI under the auspices of Compute
Canada, the Government of Ontario, the Ontario Research Fund–Research Excellence, and the University of Toronto. Research at Perimeter Institute is supported in part by the Government of Canada through the Department of Innovation, Science and Industry Canada and by the Province of Ontario through the Ministry of Colleges and Universities. This research also used resources of the National Energy Research Scientific Computing Center (NERSC), a U.S. Department of Energy Office of Science User Facility located at Lawrence Berkeley National Laboratory, operated under Contract No. DE-AC02-05CH11231 using NERSC award HEP-ERCAPmp107.

MM, AL acknowledge support from NASA grant 21-ATP21-0145. BDS, FJQ, BB, IAC, GSF, NM, DH acknowledge support from the European Research Council (ERC) under the European Union’s Horizon 2020 research and innovation programme (Grant agreement No. 851274). BDS further acknowledges support from an STFC Ernest Rutherford Fellowship. EC, BB, IH, HTJ acknowledge support from the European Research Council (ERC) under the European Union’s Horizon 2020 research and innovation programme (Grant agreement No. 849169). JCH acknowledges support from NSF grant AST-2108536, NASA grants 21-ATP21-0129 and 22-ADAP22-0145, DOE grant DE-SC00233966, the Sloan Foundation, and the Simons Foundation. CS acknowledges support from the Agencia Nacional de Investigaci\'on y Desarrollo (ANID) through FONDECYT grant no.\ 11191125 and BASAL project FB210003. RD acknowledges support from ANID BASAL project FB210003. ADH acknowledges support from the Sutton Family Chair in Science, Christianity and Cultures and from the Faculty of Arts and Science, University of Toronto. JD, ZA and ES acknowledge support from NSF grant AST-2108126. KM acknowledges support from the National Research Foundation of South Africa. AM and NS acknowledge support from NSF award number AST-1907657. IAC acknowledges support from Fundaci\'on Mauricio y Carlota Botton. LP acknowledges support from the Misrahi and Wilkinson funds. MHi acknowledges support from the National Research Foundation of South Africa (grant no. 137975). SN acknowledges support from a grant from the Simons Foundation (CCA 918271, PBL). CHC acknowledges FONDECYT Postdoc fellowship 322025. AC acknowledges support from the STFC (grant numbers ST/N000927/1, ST/S000623/1 and ST/X006387/1). RD acknowledges support from the NSF Graduate Research Fellowship Program under Grant No.\ DGE-2039656. OD acknowledges support from SNSF Eccellenza Professorial Fellowship (No. 186879). OD acknowledges support from SNSF Eccellenza Professorial Fellowship (No. 186879). CS acknowledges support from the Agencia Nacional de Investigaci\'on y Desarrollo (ANID) through FONDECYT grant no.\ 11191125 and BASAL project FB210003. TN acknowledges support from JSPS KAKENHI (Grant No.\ JP20H05859 and No.\ JP22K03682) and World Premier International Research Center Initiative (WPI), MEXT, Japan. AvE acknowledges support from NASA grants 22-ADAP22-0149 and 22-ADAP22-0150.

\bibliography{msm}
\bibliographystyle{aasjournal}

\appendix

\section{Lensing (four-point) likelihood and theory}\label{app:like}

In this appendix, we describe in more detail the components of our lensing likelihood. We approximate this as being Gaussian in the bandpowers of the estimated lensing power spectrum $\hat{C}^{\kappa\kappa}_{L_b}$:
\begin{equation}
    -2\ln{\mathcal{L}}=\sum_{bb^\prime}\big[\hat{C}^{\kappa\kappa}_{L_b}-{C}^{\kappa\kappa}_{L_b}(\boldsymbol{\theta})\big]{\mathbb{C}}^{-1}_{bb^\prime}\big[\hat{C}^{\kappa\kappa}_{L_{b^\prime}}-{C}^{\kappa\kappa}_{L_{b^\prime}}(\boldsymbol{\theta})\big],
\end{equation}
where  $C^{\kappa\kappa}_{L_b}$ is the theory lensing convergence power spectrum evaluated with cosmological parameters $\boldsymbol{\theta}$ and ${\mathbb{C}}_{bb^\prime}$ is the baseline covariance matrix for the binned spectrum, obtained from realistic sky simulations and detailed in \cite{Qu23}. When combining the lensing likelihood with that for the CMB anisotropy power spectra, we ignore the covariance between the measured lensing and anisotropy spectra, as these are negligible for DR6 noise sensitivities~\citep{Schmittfull_2013, Peloton_2017}. 

The reconstructed CMB lensing power spectrum depends on the four-point function of the CMB fields, and so, quadratically on the CMB anisotropy power spectra. We normalize the estimated lensing power spectrum with a fiducial choice of CMB power spectra, but account for the cosmology dependence of the true normalization (and of one of the bias corrections) in the likelihood analysis. For joint constraints with CMB anisotropy spectra, we correct the normalization at each point in parameter space as discussed below. For cosmology runs that do not include information from the primary CMB, we effectively marginalize over realizations of the CMB power spectrum in the normalization, informed by current constraints.  Specifically, we obtain 1000 posterior samples from the ACT DR4 + Planck primary CMB chains and propagate these to the covariance matrix as described in Appendix B of \cite{Qu23}. This step is done consistently to both the ACT and the \NPIPE\ parts of the covariance matrix.

A fiducial cosmology $\boldsymbol{\theta}_0$ is assumed in various steps of the lensing measurement. This includes the calculation of the normalization $\mathcal{R}^{-1}_L$ and $N^1_L$ bias. To account for the dependence on $\boldsymbol{\theta}_0$, the theoretical lensing power spectrum at each sampled point $\boldsymbol{\theta}$ needs to be corrected as
\begin{equation}
    {C}^{\kappa\kappa,\mathrm{th}}_{L_b}(\boldsymbol{\theta})=\frac{[\mathcal{R}^{-1}_{L_b}(\boldsymbol{\theta}_0)]^2}{[\mathcal{R}^{-1}_{L_b}(\boldsymbol{\theta})]^2}C^{\kappa\kappa}_{L_b}(\boldsymbol{\theta})-N^1_{L_b}(\boldsymbol{\theta}_0)+N^1_{L_b}(\boldsymbol{\theta}).
\end{equation}
Fully calculating the above for each point in the sampled parameter space is unfeasible, and hence we follow the approach of \cite{1502.01591,1807.06210,1611.09753} and forward model the linearized corrections to the theory spectrum due to the parameter deviations from the fiducial cosmology. For small deviations, expanding the normalization and $N^1_{L_b}$ around $\boldsymbol{\theta}$ leads to 
\begin{equation}
    {C}^{\kappa\kappa,\mathrm{th}}_{L_b}(\boldsymbol{\theta})\approx
    C^{\kappa\kappa}_{L_b}(\boldsymbol{\theta})+2\frac{d \ln{\mathcal{R}_{L_b}(\boldsymbol{\theta}_0)}}{d C^j_{\ell^{\prime}}}\Big[C^j_{\ell^{\prime}}(\boldsymbol{\theta})-C^j_{\ell^{\prime}}(\boldsymbol{\theta}_0)\Big]C^{\kappa\kappa}_{L_b}(\boldsymbol{\theta}_0)+\frac{d{N}^1_{L_b}}{d{C}^j_{\ell^{\prime}}}\Big[C^j_{\ell^{\prime}}(\boldsymbol{\theta})-C^j_{\ell^{\prime}}(\boldsymbol{\theta}_0)\Big]+\frac{d{N}^1_{L_b}}{d{C}^{\kappa\kappa}_{L^{\prime}_b}}\Big[{C}^{\kappa\kappa}_{L^{\prime}_b}(\boldsymbol{\theta})-{C}^{\kappa\kappa}_{L^{\prime}_b}(\boldsymbol{\theta}_0)\Big] .
\end{equation}

The corrections involving changes in the CMB 2-point spectra (second and third terms on the RHS above) are not included for runs that combine with the ACT DR4 CMB 2-point likelihood. This is motivated following Appendix B of \cite{Qu23} where it is noted that the ACT DR6 maps used in this work are calibrated against \Planck~ maps, and the calibration uncertainty in this process is significantly smaller than that for the maps involved in the ACT DR4 2-point CMB cosmology analysis. We have checked that an MCMC run that includes the above terms returns parameter constraints similar to those when the corrections are excluded, if the chain point spectra $C^j_{\ell^{\prime}}(\boldsymbol{\theta})$ are rescaled through calibration of $C^{TT}_{\ell^{\prime}}(\boldsymbol{\theta})$  against \Planck~ in the multipole range $1000 < \ell < 2000$. This procedure closely approximates what was done for the ACT DR6 data. For theory predictions, we use the Einstein--Boltzmann code \href{https://camb.info}{\texttt{CAMB}}~\citep{CAMB} (version v1.3.6) with sufficiently high accuracy $\text{\texttt{lmax}}=4000$, $\text{\texttt{lens\_margin}}=1250$, $\text{\texttt{lens\_potential\_accuracy}}=4$, $\text{\texttt{AccuracyBoost}}=1$, $\text{\texttt{lSampleBoost}}=1$, and $\text{\texttt{lAccuracyBoost}}=1$. While these are lower than recommended in \cite{McCarthyHM} -- see Appendix~A of \citealt{2109.04451} for importance to current-generation CMB surveys -- the evaluation time is significantly lower while being of sufficient accuracy given the precision of our measurement. We use the \texttt{mead2016} non-linear matter power spectrum prescription~\citep{Mead2015,Mead2016} with the default parameters \texttt{HMCode\_A\_baryon} $=3.13$ and \texttt{HMCode\_eta\_baryon}$=0.603$. Since our measurement mainly probes linear scales, this choice and baryonic feedback effects do not matter at current sensitivities, which can be explicitly seen in our analysis in Figure \ref{fig:sigma8_data}. For runs that include the ACT DR4 CMB anisotropy likelihood (see Appendix~\ref{app:cmb}), power spectra are calculated out to $\text{\texttt{lmax}}=7000$. We have confirmed that $\chi^2$ values from this likelihood only differ by 0.04\% when using accuracy settings from \cite{McCarthyHM} and so we do not use higher accuracy settings for ACT DR4.

We perform our MCMC inference using the \texttt{Cobaya} package \citep{2005.05290} with the Metropolis--Hastings (MH) sampler with adaptive covariance learning, and run our chains until the Gelman-Rubin criterion \citep{10.1214/ss/1177011136} for chain variances falls below $R-1=0.01$, except in cases where the curvature density is varied, where we only require a threshold of $R-1=0.02$.

\section{CMB anisotropy (two-point) likelihoods}\label{app:cmb}

While our baseline constraint on structure growth only uses the ACT (and in some cases \Planck) gravitational lensing reconstruction measurements in this work, we sometimes use information from the primary CMB anisotropies themselves either for comparison or in combination with the lensing measurement. CMB experiments like \Planck\ and ACT produce maps of the temperature (T) and polarization anisotropies (E-mode and B-mode). The angular power spectra of these maps ($TT,TE,EE$) provide information mainly on the primary anisotropies of the CMB, which depend on the early universe (redshifts $z>1100$). They, however, also are screened by reionization ($z\simeq 8$) and therefore have a dependence on the optical depth to that epoch $\tau$, and pick up secondary anisotropies like lensing. Reionization produces a suppression of the power spectra (as well as enhanced low-$\ell$ polarization) and the lensing effect induces smearing of the CMB acoustic peaks and a transfer of power from large to small scales. While the anisotropy power spectra measurements still mainly provide an early-universe extrapolation of late-time parameters like $\sigma_8$, in some cases we marginalize over an $A_{\rm lens}$ parameter that frees up the amplitude of the lensing-induced peak smearing. This isolates the early-universe information so as to allow comparison with the late-time CMB lensing reconstruction (through the CMB four-point function) and with galaxy lensing.

As our baseline for CMB anisotropies, we use data measuring the two-point function from \Planck. For the low-$\ell$ temperature component, we use the likelihood at $\ell<30$ derived from the PR3 maps \citep{planck_like:2018}. For the high-$\ell$ temperature and polarization, we use the likelihood for $TT, TE, EE$ presented in \cite{rosenberg/etal:2022}, derived from the \NPIPE\ maps~\citep{2007.04997} using the \texttt{CamSpec} likelihood. This gives consistent results to PR3 \citep{planck_like:2018,efstathiou/gratton:2021}, with around $10\%$ more constraining power. 

To include information from \Planck's large-scale polarization data that constrains the optical depth to reionization, we use the likelihood estimated in \cite{pagano/etal:2020} from the \texttt{Sroll2} maps, sampling from the \texttt{Sroll} likelihood released with PR3 but updating the data with \href{https://web.fe.infn.it/~pagano/low_ell_datasets/sroll2/}{\texttt{Sroll2}}.

We also form a second independent combination of two-point function data by combining the {\it WMAP} 9-year likelihood with the ACT DR4 likelihood, for $TT, TE, EE$. For {\it WMAP} we use the python implementation of the $\ell>23$ likelihood, \href{https://github.com/HTJense/pyWMAP}{\texttt{pyWMAP}}. For ACT we use the DR4 foreground-marginalized \href{https://github.com/ACTCollaboration/pyactlike}{\texttt{pyactlite}} likelihood software. In this case, we discard the large-scale {\it WMAP} polarization data, keeping the information on the optical depth from the \Planck\ \texttt{Sroll2} likelihood. In some cases we also test the effect of using the \Planck~ PR3 high-$\ell$ likelihood in place of the \NPIPE\ likelihood, and of approximating the optical depth with a Gaussian distribution with mean and error shifted compared to the \texttt{Sroll2} measurement.

\section{Re-analysis of galaxy weak lensing}\label{app:gal}

\begin{figure}
    \centering
    \includegraphics[width=0.7\columnwidth]{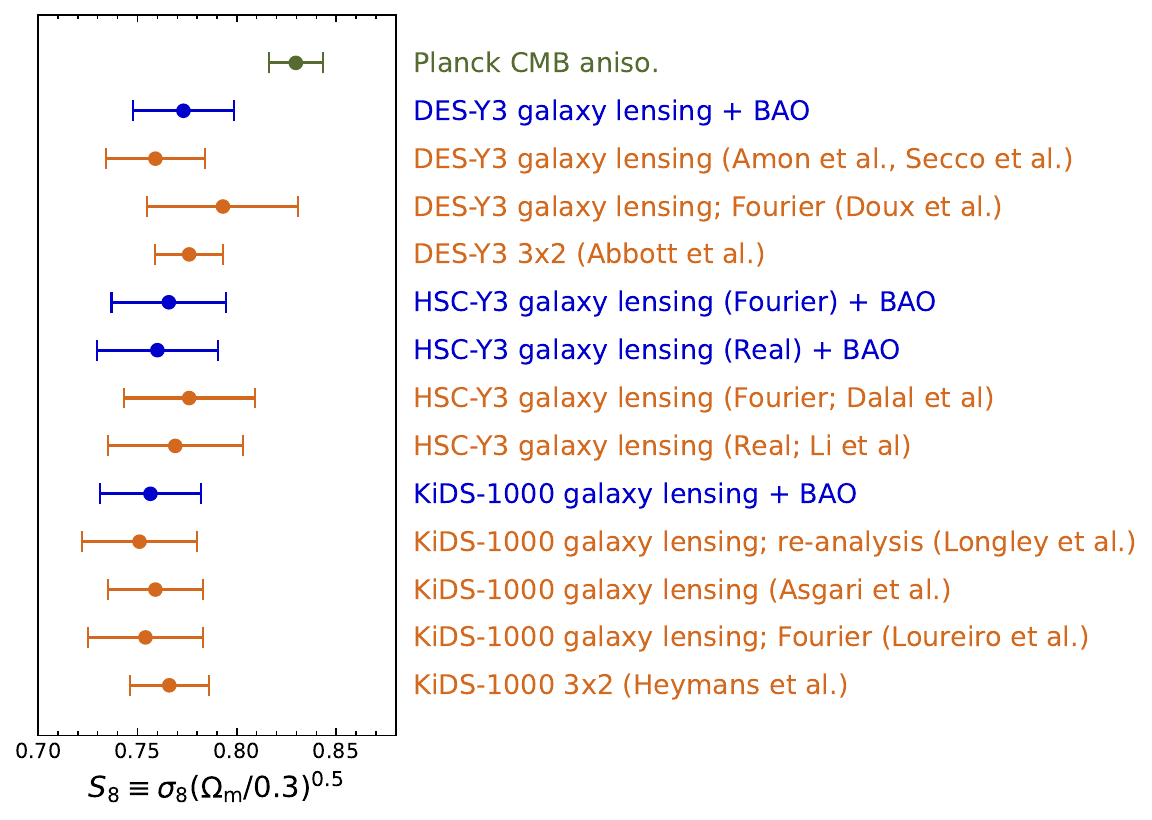}
    \caption{Comparison of $\seightg$ constraints from a consistent re-analysis of galaxy lensing (blue) with results from the literature (orange). }
    \label{fig:S8_comp}
\end{figure}

We perform our galaxy weak lensing analysis (cosmic shear) parameter inference using the \texttt{CosmoSIS} \citep{1409.3409} framework. To facilitate a consistent comparison, the re-analysis here departs from the published works from KiDS and DES in the following ways: (a) we choose the cosmological parameterization from Table \ref{tab:params} (i.e., we sample in $\loga$, $\thetamc$, $\omch$, $\ombh$ instead of $A_{\text{s}}$, $\Omega_{\text{c}}$, $\Omega_{\text{b}}$, $H_0$); (b) we choose the priors from Table \ref{tab:params}, most notably a broader prior on $H_0$ and a sharper prior on $n_{\text{s}}$; (c) we have minor differences in the version and accuracy of the CAMB Boltzmann code (see Appendix~\ref{app:like}); and (d) we sample using Metropolis--Hastings (MH) instead of a nested sampler. These choices match those from \Planck\ analyses \citep{1502.01591,1807.06210}. For the MH sampling, we use adaptive covariance learning through an interface with the \texttt{Cobaya} package \citep{2005.05290} and run our chains until the Gelman--Rubin criterion \citep{10.1214/ss/1177011136} for chain variances falls below $R-1=0.05$. We use the  \href{https://github.com/mraveri/tensiometer/}{\texttt{tensiometer}} package \citep{2105.03324,1912.04880} to load \texttt{CosmoSIS} outputs into \texttt{getdist} \citep{1910.13970}; the latter is used throughout this work to obtain marginalized one- and two-dimensional densities from MCMC samples. All reported tensions in this work use a Gaussian metric, i.e., the difference in the mean of the marginalized posteriors divided by the quadrature sum of the 68\% confidence limits for the parameter of interest. The HSC re-analyses shown here were provided by the HSC team. They were run with the same priors and parameterization as above, with the same combination of galaxy BAO, but differ in the Boltzmann codes and sampling techniques.
\begin{figure*}
    \centering
    \includegraphics[width=0.55\textwidth]{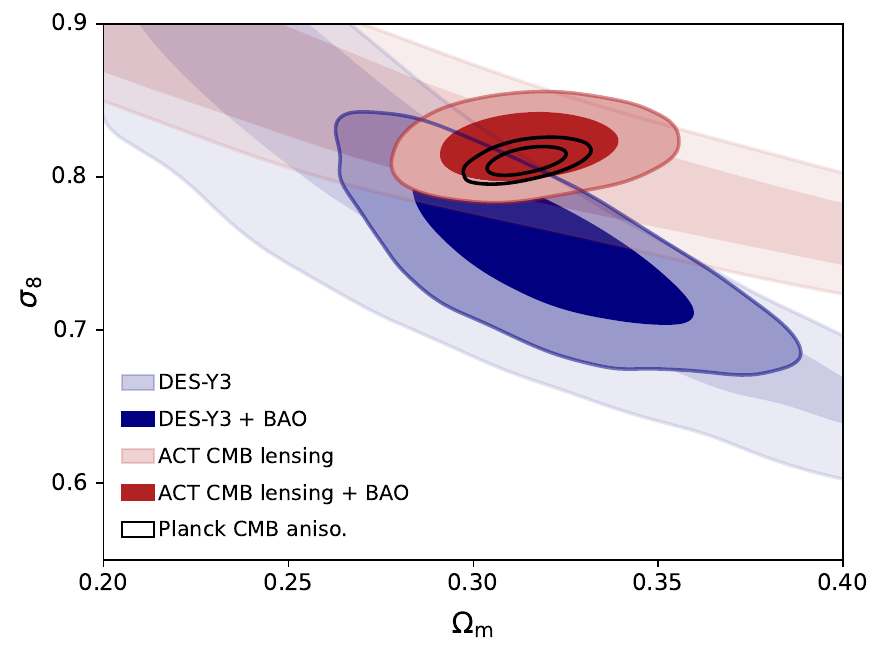}
    \caption{Constraints in the $\sigma_8-\omegam$ plane when combining ACT CMB lensing (red) or DES galaxy lensing (blue) with galaxy BAO. The posteriors in the absence of BAO are shown in lighter shades and are constrained well roughly along the $\seightc$ and $\seightg$ directions for CMB and galaxy lensing, respectively.  }
    \label{fig:s8ombao_comp}
\end{figure*}
\begin{figure*}
    \centering
    \includegraphics[width=0.49\textwidth]{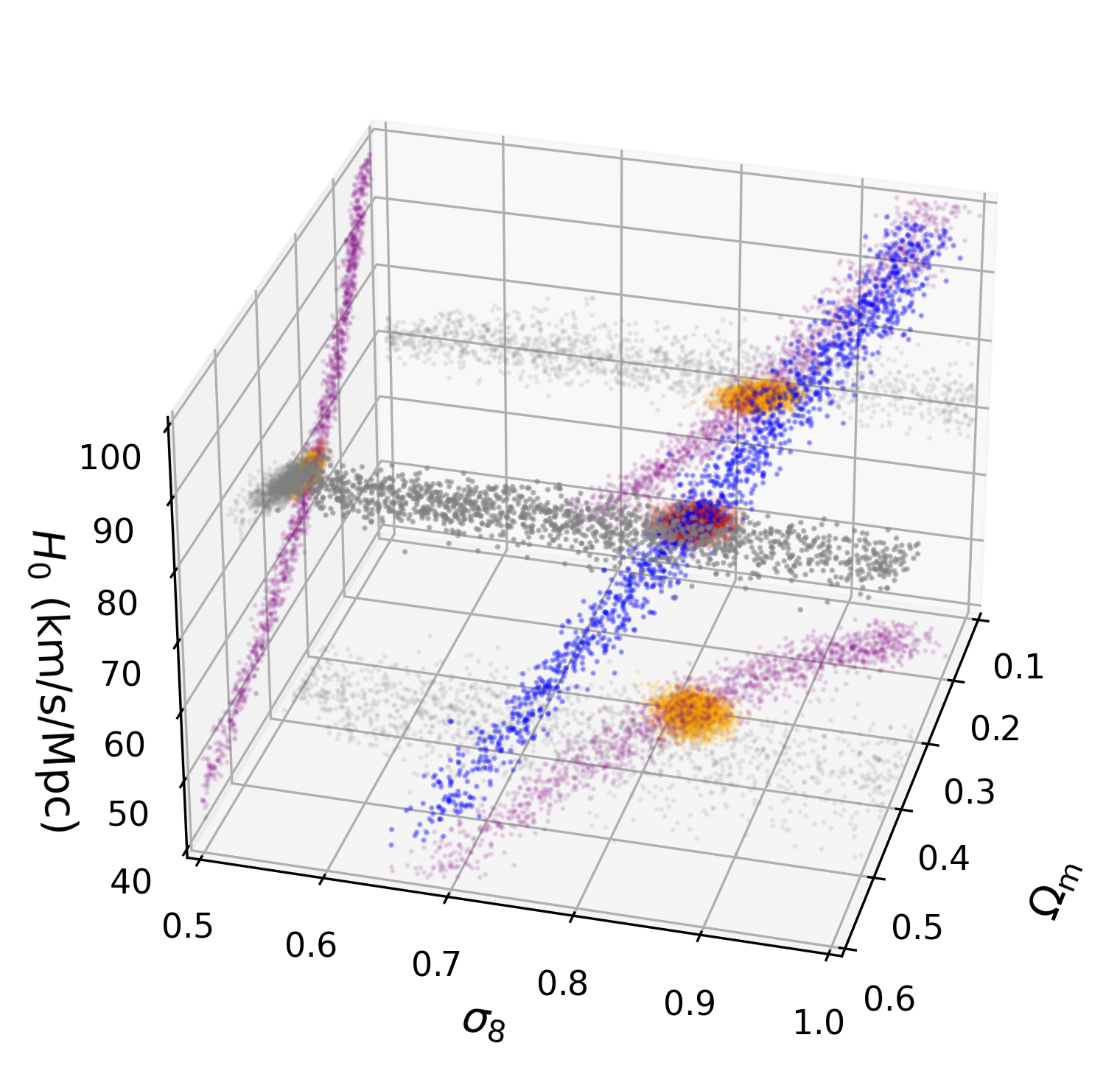}
    \includegraphics[width=0.49\textwidth]{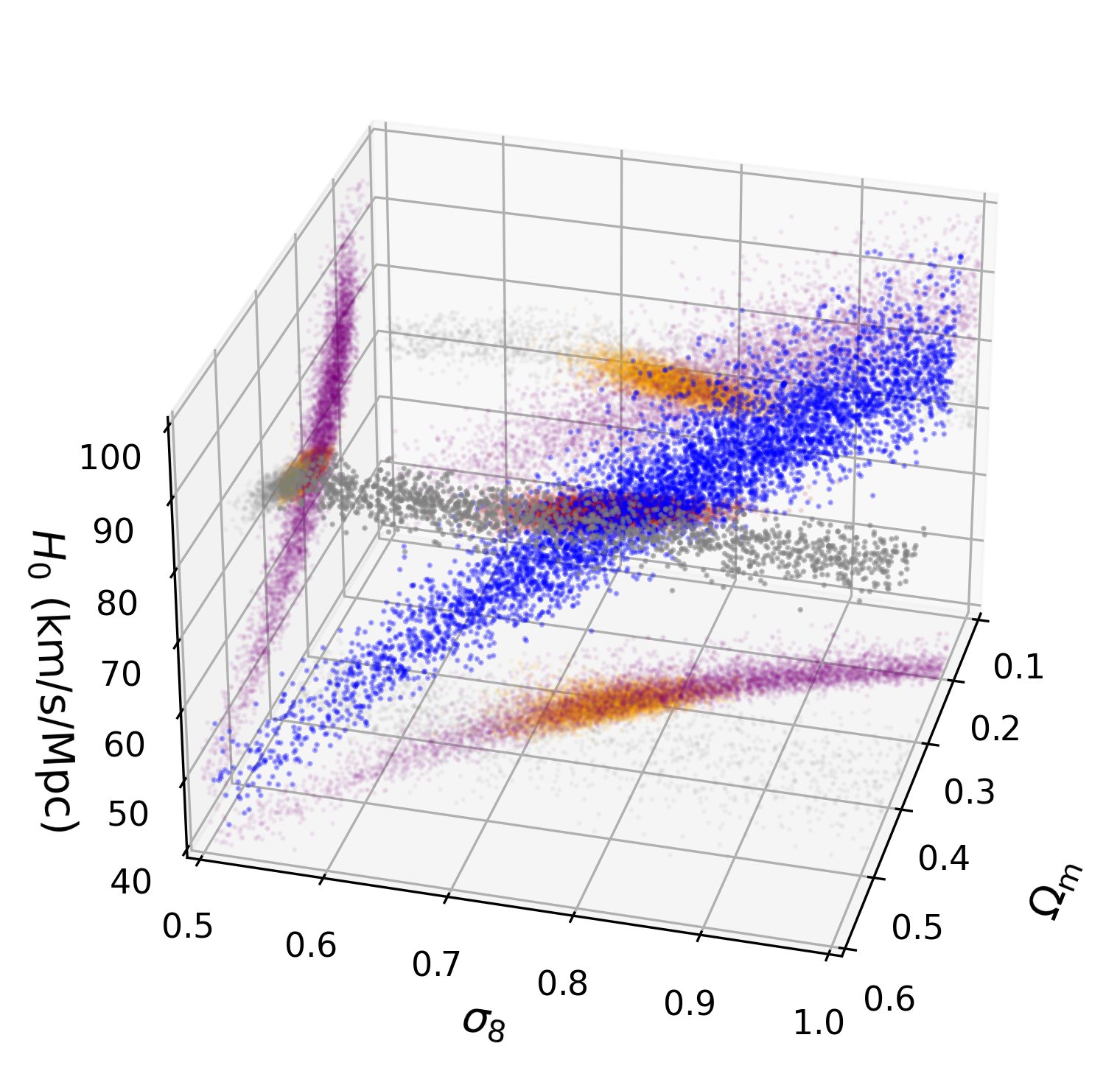}
    \caption{Distribution of MCMC samples for weak lensing in the $\sigma_8$-$\omegam$-$H_0$ space. {\it CMB lensing (left)}: Galaxy BAO samples are shown in gray; their density does not depend on $\sigma_8$. Due to the large range of scales probed by CMB lensing, shown here for ACT DR6 (blue in 3d; purple in projection), they form a line in this space. The intersection of ACT CMB lensing with BAO (red in 3d; orange in projection) provides a tight constraint on $\sigma_8$. {\it Galaxy lensing (right)}: In contrast, the galaxy weak lensing samples define a surface due to their not probing the large-scale regime, shown here for DES-Y3 for illustration (blue in 3d; purple in projection). The intersection with BAO (red in 3d; orange in projection) provides weaker constraints on $\sigma_8$.  }
    \label{fig:tubes}
\end{figure*}

In Figure \ref{fig:S8_comp}, we compare the constraints from the re-analysis (blue) with those in the literature (orange). The ``galaxy lensing'' constraints all only include cosmic shear measurements, whereas ``$\threecrosstwo$'' measurements also include galaxy clustering and galaxy-galaxy lensing. Our DES-Y3 re-analysis constraints on $\seightg$ that include BAO are in agreement with those from the DES-Y3 galaxy lensing-alone analysis in \cite{amon22} and \cite{2105.13544}, the Fourier variant of the former \citep{2203.07128}, as well as the DES-Y3 $\threecrosstwo$ analysis in \cite{2105.13549}. Similarly, our KiDS-1000 re-analysis constraints that include BAO are in agreement with those from the galaxy lensing re-analysis in \cite{2208.07179} (whose framework we follow, including for scale-cuts), the galaxy lensing analysis by the KiDS collaboration \citep{2007.15633} and its Fourier variant \citep{2110.06947}, as well as the $\threecrosstwo$ analysis by the KiDS collaboration \citep{2007.15632}. The HSC-Y3 results are consistent with those from \cite{HSCY3Fourier} and \cite{HSCY3Real}. Apart from the differences outlined above (including our choices of priors), it should be noted that some of the constraints reported in the literature do not use the marginalized mean and standard error as we do, but might report quantities such as the  multivariate maximum posterior (MAP) and its credible interval calculated using its projected joint highest posterior density (PJ-HPD), e.g. \cite{2007.15633}.

\section{Parameter dependence of CMB and galaxy lensing}\label{app:tubes}

CMB lensing constraints on parameters arise from two different ranges of scales.  First, on small scales, the CMB lensing power spectrum primarily probes the high-$k$ power-law tail of the matter power spectrum in projection; this implies that CMB lensing parameter constraints can be well approximated by a parameter combination $\sigma_8^\alpha \Omega_m^{\beta} h^\gamma$, where $\alpha, \beta, \gamma$ are constants \citep{1502.01591, 2007.04007}. On the other hand, much of our CMB lensing power spectrum constraining power also arises from intermediate and large scales, where, due to projection of the matter power spectrum near the peak, the lensing spectrum deviates from this high-$L$ power law, providing a different sensitivity to the matter-radiation equality multipole $L_{eq} \sim (\Omega_{\text{m}}^{0.6} h)$. Therefore, considering the three-dimensional $\sigma_8, \Omega_{\text{m}}, h$ parameter space, the two constraints arising from CMB lensing power spectrum constraints define two surfaces; their intersection implies that the CMB lensing power spectrum constraints define a line in this space. Now we can easily explain why the constraint on $\sigma_8$ when combining with BAO (as seen in Figure \ref{fig:s8ombao_comp}) is so tight: BAO defines another surface in this space, so that the intersection of the BAO and lensing constraints is a point (or a small region in parameter space; see left panel of Figure \ref{fig:tubes}).

In contrast, galaxy lensing generally does not probe the large-scale regime of scales approaching the matter power spectrum peak; effectively, it only provides one small-scale constraint within the $\sigma_8, \Omega_{\text{m}}, h$  space, defining a single surface. Adding the BAO data, which defines a different surface, the intersection gives a line-shaped constraint (instead of a point as for the CMB lensing and BAO combination; see right panel of Figure \ref{fig:tubes}). This explains why the $\sigma_8$ constraint is much broader and shows a significant degeneracy with the matter density.

We have verified this explanation with a simple exercise: we perform an analysis on mock CMB lensing data, artificially adjusting the errors to vary the scales from which the information originates, while holding the total signal-to-noise constant. When we shift the mock CMB lensing information only to arise from small scales, $L>2000$, the shapes of the parameter constraint contours and the constraints on $\sigma_8$ show a close resemblance to the constraints from the combination of galaxy weak lensing and BAO.

\end{document}